\documentclass[review]{article}

\usepackage{authblk}
\usepackage{dcolumn}
\usepackage{bm}

\usepackage[english]{babel}
\usepackage[utf8]{inputenc}
\usepackage[T1]{fontenc}
\usepackage{mathptmx}
\usepackage{etoolbox}

\usepackage{graphicx} 
\usepackage{wrapfig}
\usepackage{subcaption}
\usepackage{amsmath} 
\usepackage{amsthm}
\usepackage{amssymb} 
\usepackage{xcolor}
\usepackage{multirow}
\usepackage{bbm}
\usepackage{mathtools}
\usepackage{makecell}

\renewcommand{\selectlanguage}[1]{}

\begin{document}


\title{Quantification of the cascading tipping probability from the AMOC to the Amazon rainforest with a rare-event algorithm} 



\author[1]{Valérian Jacques-Dumas\thanks{v.s.jacques-dumas@uu.nl}}

\author[1,2]{Henk A. Dijkstra}
\affil[1]{Institute for Marine and Atmospheric research Utrecht, Department of Physics,
Utrecht University, Princetonplein 5, 3584 CC Utrecht, Netherlands}
\affil[2]{Center for Complex Systems Studies, Department of Physics,
Utrecht University, Leuvenlaan 4,
3584 CE, Utrecht, Netherlands}


\date{\today}

\maketitle

\begin{abstract}
The Amazon rainforest and the AMOC are considered to be tipping elements: they are important components of the Earth system, but may collapse under climate change.
Moreover, an AMOC collapse may favor the transition of the rainforest to a degraded forest by influencing the precipitation patterns over the Amazon.
This phenomenon is known as tipping cascade and better understanding it is key to anticipating the impact of tipping events.
Here, we investigate in a coupled conceptual AMOC-Amazon model the probability that an AMOC weakening affects tree cover loss in two regions of the rainforest.
To get more insight into the mechanisms behind the tipping cascade, we also analyze the dynamics of both systems and their evolution during the Amazon transition.
Namely, we track the transition probability and the transition time of the Amazon, and reconstruct the distribution of AMOC strength at every stage of this transition.
These tasks require a large ensemble simulation, containing in particular a large number of transitions.
Since such events may be too rare to be sampled by direct numerical simulation, the collapse of both systems is studied using TAMS, a ``rare-event'' algorithm designed to efficiently sample rare transitions.
We find that, in the northwest of Brazil, a transition of the Amazon rainforest to a degraded forest within $200$ years is very unlikely.
However, in this region, such transition can only occur after an AMOC collapse, which would have a large drying effect that favors the development of extreme wildfires.
\end{abstract}

{\bf
    The Amazon rainforest and the Atlantic Meridional Overturning Circulation (AMOC) are thought to be tipping elements: they are large-scale systems playing a major role in the Earth system, but they may abruptly collapse under climate change.
    An AMOC collapse would have important consequences on the climate system, among which important perturbations of the precipitation patterns over the Amazon.
    Such a change might destabilize the rainforest and favor an abrupt decrease in its tree cover, leading to its transition to a degraded forest.
    This phenomenon is called a tipping cascade.
    Assessing its likelihood and, more importantly, understanding its underlying mechanisms and the interactions of tipping elements, are key to better preparing for its impacts.
    Here, we set out to quantify the cascading tipping probability from the AMOC to the Amazon in a conceptual model.
    But, although informative, such indicator is not sufficient to obtain insight into the coupled dynamics of the AMOC and the Amazon.
    To better understand their interaction, we have to track other dynamical quantities at different stages of the Amazon transition.
	In this way, we can precisely reconstruct the influence of the AMOC over the Amazon at any point during its transition from a rainforest to a degraded forest.
    For instance, it is particularly interesting to analyze the transition time of the Amazon (since different transition mechanisms may occur on different time scales) and the distribution of the AMOC strength (since certain values of AMOC strength may favor the Amazon transition).
    We carry out this statistical analysis using a so-called rare-event algorithm, which is meant to ease the sampling of rare transitions (such as an AMOC or Amazon collapse) by biasing ensemble simulations in a controlled way.
	We find that, in the northwest of Brazil, a transition of the Amazon rainforest within $200$ years is unlikely.
	But, in this region, an AMOC collapse would favor such transition by reducing precipitation and allowing the development of extreme wildfires.
}

\section{Introduction}

Tipping elements are important components of the Earth's climate system because they may undergo abrupt transitions, bringing them to a disrupted state \cite{Lenton2008,ArmstrongMcKay2022}.
We focus here on two subsystems: the Atlantic Meridional Overturning Circulation (AMOC) and the Amazon rainforest.

The AMOC plays a major role in meridional heat transport, thus influencing the climate of the Northern Hemisphere and, more generally, of the entire planet.
\cite{Stommel1961} was the first to suggest it may be prone to tipping by finding bistability in a conceptual model.
Since then, hysteresis behavior has been found across the whole hierarchy of models~\cite{Weijer2019}.
In particular, \cite{vanWesten2024a} simulated an AMOC collapse in the Community Earth System Model (CESM), a fully coupled General Circulation Model (GCM).
Van Westen, Kliphuis, and Dijkstra (2024) highlighted the possibility that the AMOC may be on tipping course using a physical indicator: the freshwater transported by the AMOC across the southern boundary of the Atlantic Ocean.
Available observations~\cite{Bryden2011,Garzoli2013} of this indicator suggest that the AMOC is indeed in such a bistable regime.
Note, however, that most models from the Coupled Model Intercomparison Project phase 6 (CMIP6) disagree on the current-day AMOC amplitude~\cite{Weijer2020} and suffer from freshwater biases~\cite{vanWesten2024b}.
As a result of these biases, they are likely too stable, which may even prevent the existence of a stable collapsed state~\cite{Mecking2017,Gent2018,Vanderborght2025}.

Tipping events can be categorized into three main types~\cite{Ashwin2012}: bifurcation-induced tipping, rate-induced tipping and noise-induced tipping.
Bifurcation-induced tipping occurs when one stable equilibrium disappears after a control parameter (representing, for instance, global warming) of the system has reached a critical threshold.
When this parameter evolves at a fast-enough rate, non-autonomous effects can trigger the tipping before the critical threshold.
Noise-induced tipping is due to stochastic jumps between stable equilibria.
When different systems interact on very different time scales, stochastic processes are a convenient way of modeling the influence of the faster system on the slower one.
For instance, atmospheric variability can be modeled as a stochastic forcing of the ocean salinity.
\cite{Slyman2023} showed that the interaction of non-autonomous and stochastic effects may increase the tipping probability, and tipping may occur earlier than with rate-induced effects alone.
Therefore, stochastic effects may play a crucial role in tipping dynamics and must be taken into account.
We focus here, for simplicity, on noise-induced tipping only, but we showed in~\cite{Jacques-Dumas2024a} that our framework works the same way for non-autonomous systems.
Under pre-industrial conditions, a purely noise-induced AMOC collapse may be very rare, when comparing its relatively short time scale to the absence of evidence that it occurred in the historical period~\cite{Soons2024}.
Van Westen \textit{et al.} (2025) showed that time-dependent anthropogenic forcing on the climate system may make the onset of such collapse more likely in the coming century but the probability of occurrence of an AMOC tipping in the near-future remains difficult to estimate.
A recent study by~\cite{Ditlevsen2023} used observational data to show that, under anthropogenic forcing, the AMOC may tip between 2037 and 2109 with $95\%$ certainty.
It has however been criticised since, because it builds on the idea of critical slowing down~\cite{Dakos2008,Boulton2014,Boers2021}, which relies on assumptions regarding noise, stationarity, and is not fit for application to observational data~\cite{benYami2024b}.
The method is known to be prone to spurious alarms when these assumptions are not fulfilled~\cite{vanWesten2024a,Zimmerman2025,Rietkerk2025}.

The Amazon rainforest plays a crucial role in the global carbon cycle, is an important reservoir of biodiversity, and cools the global climate through the evapotranspiration~\cite{IPCC2021}.
However, it has been shown that its tree cover may abruptly respond to changes in hydrological conditions and could possess multiple stable states~\cite{Hirota2011}, namely a (current) rainforest state, a state with low tree cover (tree cover comparable to that of a savanna) and a treeless state.
\cite{Flores2024} have recently pinpointed several forcing factors that may bring the Amazon out of its rainforest state, including water stress, global warming and deforestation.
Furthermore, atmospheric moisture flow~\cite{Flores2024} creates connectivity between geographically separated regions of the forest, as evaporation in a given region may influence precipitation in others.
Such connectivity may increase the forest's sensitivity to change~\cite{Scheffer2012} by acting as a positive feedback on precipitation.
Indeed, a region receiving less moisture inflow undergoes reduced rainfall, which reduces in turn evapotranspiration and moisture outflow~\cite{Zemp2017}.
The reduction in precipitation can then cascade across other regions of the Amazon and facilitate the abrupt transition of the system~\cite{Wunderling2025}.

Until now, we have presented tipping elements separately, but they are not isolated in practice.
Since they interact in the Earth system, tipping of one subsystem\,--\,called leading\,--\,might affect the dynamics of another tipping element\,--\,called following, which may then tip in turn.
This phenomenon is called a tipping cascade~\cite{Dekker2018} and has gained considerable attention in recent years (see~\cite{Wunderling2025} for an extensive review), since such a cascade may have a global impact on the Earth system.
\cite{Kriegler2009} assessed the likelihood of tipping cascades in a network of five tipping elements, and concluded that the effect of the AMOC on the Amazon rainforest is uncertain.
An AMOC weakening would likely cause a southward shift in the Intertropical Convergence Zone (ITCZ)~\cite{Jackson2015,Bellomo2023,benYami2024}, thus augmenting water stress in the northern part of the Amazon rainforest while increasing precipitation over the part of the Amazon rainforest located in the Southern Hemisphere.
But since the southern part of the Amazon rainforest is the largest contributor to rainfall generation~\cite{Staal2018}, rainfall strengthening there may stabilize the system as a whole.
Moreover, an AMOC weakening may shift the seasonal cycle, making the wet season dryer and the dry season wetter~\cite{Parsons2014,benYami2024}, which would have a still unknown effect.

For these reasons, it is crucial to better understand the relationship between both systems and to quantify their interconnection.
The most straightforward way of doing so is to compute the cascading tipping probability, i.e. the probability that AMOC tipping triggers a tipping of the Amazon rainforest.
But, although directly informative, such indicator is not sufficient to understand the connection between the dynamics of the AMOC and those of the Amazon.
Moreover, the cascading probability can only be made meaningful when backed by an analysis of the underlying cascading processes.
The estimation of the cascading probability should therefore be accompanied by a study of the state of both systems at every stage of the tipping cascade, and of the time scale of the tipping cascade.

However, such in-depth study can only be carried out using a large ensemble simulation of tipping cascades of the coupled AMOC-Amazon system.
It is then possible to track among these simulations the relevant quantities that contain information about the whole transition.
For instance, looking at the mean transition time of the Amazon to a degraded forest, may inform about the transition mechanism, since different mechanisms may unfold on different time scales.
Similarly, studying the distribution of AMOC strength at every stage of the transition of the Amazon forest may show that some values of AMOC strength favor the Amazon transition more than others.
Because of the computational cost associated with such ensemble simulation in large climate models, the in-depth quantitative analysis of tipping cascades has been until now limited to simple, conceptual models.
In particular, the cascading probability for the AMOC-Amazon coupled system has only been estimated in non-process-based models~\cite{Ciemer2021}, where tipping elements are generally assumed to obey a simple double-fold bifurcation and to be linearly coupled~\cite{Wunderling2021,Wunderling2021b,Wunderling2023}.
When the studied cascade is unlikely, or when its likelihood is difficult to estimate, the reliance on large ensemble simulations becomes even more problematic.
In this case, the cost of uncertainty comes in addition to the computational burden: it is not guaranteed that, after running out of computing resources, we have sampled even a couple of tipping cascades.
Without a sufficient amount of sampled events, any statistical analysis becomes meaningless and numerous trajectories have been simulated to no avail.

Therefore, the computational cost of large ensemble simulation calls for the use of more efficient methods to sample and analyze potentially rare events.
Rare-event algorithms were precisely developed to sample rare events much more efficiently, by biasing ensembles of trajectories in a controlled way.
Even if the event of interest happens to be more likely than expected (think of an AMOC collapse under time-dependent anthropogenic forcing), turning to such methods can be thought of as a precautionary principle.
Different types of rare-event algorithms have recently been applied to the exploration of rare and extreme events in climate systems~\cite{Ragone2018,Webber2019,Ragone2021,Jacques-Dumas2024b,Cini2024}.
Here, we focus on an algorithm called Time Adaptive Multilevel Splitting (TAMS)~\cite{Lestang2018}, which computes the probability that a system reaches a certain region of phase space before a specific time horizon.

Our main goal in this paper is to compute the cascading tipping probability from the AMOC to four different regions of the Amazon rainforest in a coupled conceptual model based on physical processes.
This probability gives insight into the cascading tipping probability from the AMOC to the Amazon rainforest and, more generally, the connection between the AMOC and the Amazon rainforest in two different regions in the north of Brazil.
The models used here are relatively simple, but unlike in previous studies, they are not a priori supposed to be in a double-fold form.
We compute transition probabilities by sampling transitions through the application of TAMS to the conceptual model.
But, more importantly, rather than just outputting a single number, TAMS simulates an ensemble of trajectories that explore the different stages of the transition.
This property makes it also particularly suited to an in-depth analysis of the interaction between the AMOC and the Amazon, by providing the necessary data to conduct it.

This paper is structured as follows.
First, we detail in Section~\ref{sec:model} the coupled conceptual model used here.
The rare-event algorithm is then described in Section~\ref{sec:tams}, along with the way it is used to estimate dynamical quantities.
We analyze in Section~\ref{sec:results} the results of the rare-event algorithm, the impact of the AMOC on the Amazon rainforest and the cascading tipping probability between both systems.
Finally, we provide in Section~\ref{sec:discussion} a discussion of these results and possible perspectives for future work.

\section{Model description}
\label{sec:model}

The goal of this study is to analyze the cascading tipping from the AMOC to the Amazon rainforest.
In particular, we are interested in computing the probability that the Amazon rainforest may transition to a degraded forest as a consequence of an AMOC collapse.
This probability estimation is performed using a rare-event algorithm, which relies on the simulation of a large ensemble of AMOC-Amazon centennial trajectories.
However, such ensemble simulation is not feasible in a General Circulation Model (GCM).
Therefore, the AMOC and the Amazon rainforest are each simulated by a stochastic conceptual model.
To make their dynamics as close to those of a GCM as possible, these models and their coupling are tuned to the Community Earth System Model (CESM, version $1.0.5$~\cite{Hurrell2013}).
CESM is a General Circulation Model (GCM), chosen because it was used by~\cite{vanWesten2024a} to simulate an AMOC collapse.
Therefore, by using the dataset from~\cite{vanWesten2024a}, we have access to the dynamics of the Amazon rainforest in a large fully-coupled model at every stage of an AMOC collapse.

The AMOC conceptual model (see Sect.~\ref{sec:amoc_model}) is initialized from a present-day-like circulation, called on-state, whose strength is tuned to the strength of the AMOC on-state in CESM.
The Amazon rainforest (see Sect.~\ref{sec:amazon_model}) is described by its zonally averaged tree cover.
The tree cover is simulated by a model based on the empirical distribution of two hydrological variables: Mean Annual Precipitation (MAP) and Maximum Cumulative Water Deficit (MCWD).
MAP and MCWD themselves are not simulated: they are computed in CESM, where their relation to the AMOC strength is also derived (see Sect.~\ref{sec:cesm}).
Values of MAP and MCWD can then be deduced from every simulated value of AMOC strength, and they are used to determine the tree cover.
So, the AMOC simulated by the AMOC conceptual model drives the evolution of the hydrological variables governing the Amazon model.
Here, the conceptual AMOC model thus plays the role of leading system, while the conceptual Amazon model plays the role of following system.

\subsection{AMOC model}
\label{sec:amoc_model}

\begin{figure}
    \centering
    \includegraphics[width=\textwidth]{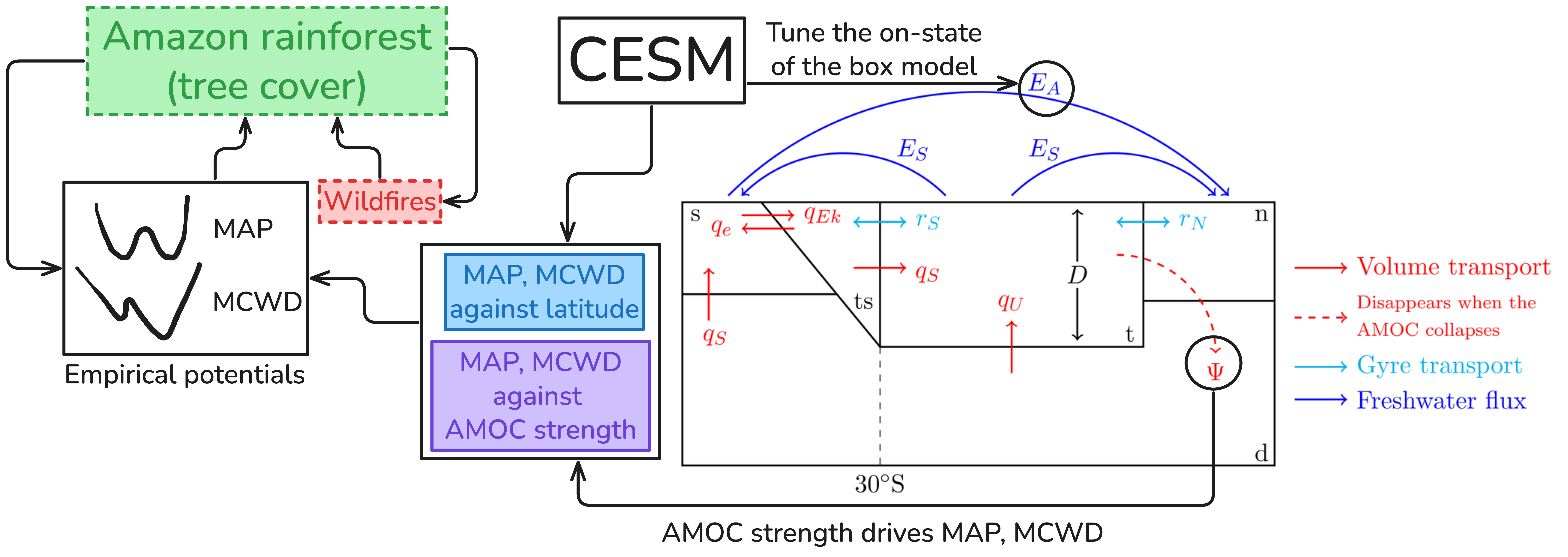}
	\caption{The AMOC conceptual model (see Sect.~\ref{sec:amoc_model}) is represented on the right. Arrows represent volume transport between boxes. Solid red arrows correspond to fluxes that are always present. The dashed red arrow represents the downwelling in the northern Atlantic, and its shutdown characterizes an AMOC collapse. The on-state of the AMOC conceptual model is tuned to that of CESM. CESM is used to derive a relationship between the AMOC strength and MAP and MCWD (see Sect.~\ref{sec:amazon_model}). The dynamics of the Amazon conceptual model are governed by the empirical potentials of MAP and MCWD, and by the stochastic process representing wildfires. The dynamics of the potentials and the fire intensity depend, in turn, on the tree cover.}
	\label{fig:coupled_model}
\end{figure}

The conceptual AMOC model used here was first introduced by~\cite{Cimatoribus2014}.
It consists of five boxes (Fig.~\ref{fig:coupled_model}): the northern Atlantic (denoted as $n$), the pycnocline (divided into two boxes, denoted as $t$ and $ts$), the deep ocean (denoted as $d$) and the Southern Ocean (denoted as $s$), which is not part of the Atlantic basin itself.
The box $ts$ lies to the south of $30^\circ$~S, between the pycnocline and the surface.
It is separated from the rest of the pycnocline box to highlight the meridional density gradient across the Atlantic basin, which is the driver of the AMOC~\cite{Cimatoribus2014}.
Moreover, the division between boxes $t$ and $ts$ allows for a better description of the transport mechanisms across the pycnocline~\cite{Cimatoribus2014}.
Water density depends on temperature and salinity, but in this model, the temperature is fixed in all boxes.
Indeed, temperature anomalies decay faster than salinity anomalies in the ocean so, in simple models, salinity is often considered more important than temperature to the long-term AMOC dynamics~\cite{Cimatoribus2014}.
The only model variables are thus the salinity of each box and the pycnocline depth $D$, so the state vector reads $\mathbf{x}= (S_t,\ S_{ts},\ S_n,\ S_s,\ S_d,\ D)$.
The volume transports between each box are represented by the red and blue arrows in Fig.~\ref{fig:coupled_model}.
Among those, the three most important transports are $\Psi$, $q_S$ and $q_U$.
The downwelling in the Northern Atlantic is represented by $\Psi$, which also corresponds to the AMOC strength.
The term $q_S$ denotes volume transport between the southern and tropical Atlantic and is the difference between the wind-driven Ekman flow ($q_{Ek}$) and the eddy-induced volume transport ($q_e$).
Finally, $q_U$ stands for the Ekman upwelling through the pycnocline.
The wind-driven subtropical gyre volume transports are simply parametrized by two constants $r_s$ and $r_n$.

The model dynamics are determined by a surface freshwater flux separated into two parts: a fixed symmetric component $E_S$ from the box $t$ to boxes $s$ and $n$, and a varying asymmetric component $\overline{E_A}$ from box $s$ to box $n$.
The symmetric component $E_S$ represents the precipitation / evaporation difference between the tropics and the subpolar boxes.
The asymmetric component $\overline{E_A}$ acts as an analog of the ``freshwater hosing flux'' implemented in large climate models.
Such a hosing flux allows pouring freshwater into the northern Atlantic while compensating it elsewhere (here in box $s$) to keep a closed salinity budget~\cite{Cimatoribus2014}.
Since water density in this model only depends on salinity, $\overline{E_A}$ determines the meridional density gradient across the Atlantic basin and acts as a control parameter governing the dynamics of the AMOC.
\cite{Castellana2019} extended the model by multiplying $\overline{E_A}$ by a stochastic term representing atmospheric variability: $E_A(t) = \overline{E_A}(1+f_n\zeta(t))$, where $\overline{E_A}$ and $f_n$ are two fixed model parameters, and $\zeta(t)$ is a white noise process with zero mean and unit variance.
Here, we only study noise-induced AMOC collapses, so we also fix the parameters $\overline{E_A}$ and $f_n$ (see the end of this section for the precise values).

When $\overline{E_A}$ belongs to the range $[0.06,0.35]$~Sv (where $1$~Sv$\equiv10^6$~m$^3\,$s$^{-1}$), the modeled AMOC possesses two coexisting stable steady states.
One of them, called the AMOC on-state and denoted as $\mathbf{x}_\mathrm{ON}$, is characterized by $\Psi>q_S>q_U>0$.
It corresponds to a present-day-like circulation, with a strong downwelling in the northern Atlantic.
The other stable state is characterized by $\Psi=0$ and $q_S<0$: the downwelling in the northern Atlantic is stopped and the circulation in the Southern Hemisphere is reversed.
The transition from the stable on-state to the stable off-state occurs on a millenial time scale~\cite{Castellana2019}.

Here, we are rather interested in simulating an AMOC shutdown within a relatively short (centennial) time scale, to analyze its relation to a transition of the Amazon rainforest.
There is, in this conceptual AMOC model, another mechanism for an AMOC shutdown, which occurs on a centennial time scale.
The downwelling in the northern Atlantic may also be shut down without a reversal of the rest of the circulation, due to large stochastic freshwater inputs through the hosing flux $E_A$.
In this situation, the AMOC strength $\Psi$ becomes zero since the downwelling does not occur anymore but stochastic perturbations are too fast to disturb the deep ocean.
Therefore, the circulation in the Southern Hemisphere is left unchanged.
Since the upwelling in the southern ocean still takes place, water accumulates in the pycnocline box, so such AMOC shutdown can only be transient and the AMOC eventually recovers.
Hence, what we call here an AMOC collapse is a transition from the steady AMOC on-state to any AMOC state whose strength is $\Psi=0$.

The steady AMOC on-state $\mathbf{x}_\mathrm{ON}$ is characterized here by the strength of its downwelling $\Psi$ and is controlled by $\overline{E_A}$.
In CESM, the strength of the AMOC on-state $\Psi_\mathrm{ON}=15.92$~Sv is defined as the average AMOC strength over the first $100$ years of the simulation by~\cite{vanWesten2024a}.
We set the value of $\overline{E_A}$ to $0.22$~Sv to match the value of $\Psi$ in the state $\mathbf{x}_\mathrm{ON}$ to that of $\Psi_\mathrm{ON}$ in CESM.
\cite{Castellana2019} determined from precipitation-evaporation records that the noise amplitude $f_n\overline{E_A}$ should be set to $0.02$~Sv, which we also follow here.
The parameters and equations of the model are given in~\cite{Castellana2019,Jacques-Dumas2023} and not repeated here.

\subsection{Amazon model}
\label{sec:amazon_model}

It has been found in many models~\cite{vanWesten2024a,Orihuela-Pinto2022,benYami2024} that an AMOC collapse would mostly affect rainfall patterns over the Amazon due to a meridional shift in the Intertropical Convergence Zone (ITCZ).
As a part of the Hadley cell, the ITCZ has a strong meridional component; hence, we perform a zonal averaging~\cite{Mamalakis2021} and model the Amazon rainforest along latitude only, denoted as $\theta$.
Moreover, we describe the Amazon rainforest using its tree cover $T$ only, as was done, for instance, in~\cite{Staal2015}.
Therefore, our model of the Amazon rainforest consists of a partial differential equation describing the tree cover $T$ as a function of time $t$ and latitude $\theta$, depending on MAP and MCWD (computed in the AMOC model):
\begin{equation}
\label{eq:amazon_model}
	\frac{\partial T}{\partial t} = D\frac{\partial^2T}{\partial \theta^2} -\frac{1}{2}\left(\nabla U_{MAP} + \nabla U_{MCWD}\right) + \sigma f(T,MAP)\eta^\alpha,
\end{equation}
where all terms are described below and all parameters are summarized in Table~\ref{tab:params}.

The first term of Eq.~\ref{eq:amazon_model} is a spatial diffusion term~\cite{Wuyts2019}, where $D$ denotes the meridional diffusion coefficient.
The second term describes the dependence of the tree cover $T$ on two hydrological variables: Mean Annual Precipitation (MAP), Maximum Cumulative Water Deficit (MCWD).
The functions $U_{MAP}$ and $U_{MCWD}$ correspond respectively to the MAP and MCWD potentials, based on the empirical distributions presented in~\cite{Flores2024} (see Extended Data Fig.~1 therein) of MAP against tree cover and MCWD against tree cover.
These distributions were obtained from the MODIS~\cite{modis} dataset (Moderate Resolution Imaging Spectroradiometer), containing local measurements of tree cover, MAP and MCWD realized in the Amazon basin.
\cite{Flores2024} showed that, when plotted against MAP or MCWD, $T$ possesses three stable states: a rainforest state (around $80\%$ tree cover), a low tree cover state (around $30\%$ tree cover) and a treeless state.
We used that same dataset to derive the combined MAP-MCWD potential, as detailed in the next subsection (Sect.~\ref{sec:potentials}).

Finally, the third term of Eq.~\ref{eq:amazon_model} is a noise process modeling changes in $T$ occurring on a faster time scale than the numerical time step (typically a few days), in particular wildfires.
The $\alpha-$stable noise process~\cite{Zheng2025} itself is denoted as $\eta^\alpha$ and its amplitude is denoted as $\sigma$ (see Sect.~\ref{sec:fire_term} for details).
The function $f(T,MAP)$ denotes the physical fire amplitude depending on tree cover and MAP.

The dynamics of the model result from the interplay between vegetation diffusion and the most stable tree cover value allowed by the combined potential of MAP and MCWD.
The presence of noise occasionally perturbs the system on a faster time scale.
At the domain boundaries, we apply von Neumann (no-flux) boundary conditions.
We solve the equation for tree cover (Eq.~\ref{eq:amazon_model}) using a discrete time step $\mathrm{d}t=0.05$ years and a spatial step $\mathrm{d}\theta=0.04^\circ$.
To ensure numerical stability of the solution independently of these parameters, Eq.~\ref{eq:amazon_model} is integrated using the Backwards Euler scheme.

\begin{table}
	\centering
	\begin{tabular}{c|c|c|c}
		Symbol & Parameter & Value & Reference\\
		\hline
		$D$ & Horizontal diffusion coefficient & $0.1$~$\mathrm{km}^{2}.\mathrm{year}^{-1}$ & \cite{Wuyts2019}\\
		\hline
		$\sigma$ & Noise amplitude & $0.05$ & \cite{Zheng2025}\\
		\hline
		$\gamma$ & Power in fire-induced mortality term & $6$ & \cite{Staal2015}\\
		\hline
		$\beta$ & Power in continuity function & $6$ & \cite{Staal2015}\\
		\hline
		$\delta$ & Power in soil moisture index function & $4$ & \cite{Staal2015}\\
		\hline
		$h_I$ & \makecell{Half saturation of the\\fire-induced mortality term} & $0.15$ & \cite{Staal2015}\\
		\hline
		$h_C$ & \makecell{Half saturation of\\grass (non-forest) cover continuity} & $0.57$ (fractional tree cover) & \cite{Staal2015}\\
		\hline
		$h_{SMI}$ & Half saturation of the soil moisture index & $1800$~$\mathrm{mm}.\mathrm{year}^{-1}$ & \cite{Staal2015}
	\end{tabular}
	\caption{The Amazon model's parameters with their value and the reference where we found these values.}
	\label{tab:params}
\end{table}

\subsubsection{Empirical potentials}
\label{sec:potentials}

Here, we describe the computation of the potentials $U_{MAP}$ and $U_{MCWD}$ used in the second term of Eq.~\ref{eq:amazon_model}.

MAP simply consists of the yearly averaged precipitation on every point of the Amazon rainforest and represents the mean ``wetness'' of this location.
MCWD~\cite{Aragao2007,Malhi2009} describes the accumulated water stress month after month by combining the intensity and duration of the dry season, which makes it a richer indicator than the mere dry season length.
MCWD is defined as the yearly maximum of Climatological Water Deficit (CWD, in mm/month), corresponding to the monthly accumulated difference between monthly precipitation ($P_k$, where $k\in[1,12]$ numbers the months) and expected evapotranspiration (constant arbitrary value of $E=100$~mm/month~\cite{Aragao2007}).
Each month $k$, CWD is updated by adding $P_k-E$ to the pre-existing water deficit from the former month.
As a deficit, CWD is always negative and bounded above by zero when the soil is saturated, so additional precipitation cannot be absorbed anymore.
Computation of the CWD starts at the wettest month of the year, assuming that the soil is saturated.
\begin{equation}
\label{eq:mcwd}
	\begin{aligned}
		&\mathrm{CWD}_0 = 0 \\
		&\mathrm{CWD}_k = \min(0, \mathrm{CWD}_{k-1} + \mathrm{P}_k - \mathrm{E}), \forall k\in[1,12] \\
		&\mathrm{MCWD} = {\max}(\mathrm{CWD}_1,\ldots,\mathrm{CWD}_{12})
	\end{aligned}
\end{equation}
MAP and MCWD are not completely uncorrelated, but we assume that the average precipitation and the characteristics of the dry season are complementary enough to allow adding both potentials in Eq.~\ref{eq:amazon_model}.
Moreover, we give each potential an equal weight in the sum, as it is the most neutral assumption on the influence of MAP and MCWD on the tree cover.

\begin{figure}
	\centering
	\includegraphics[width=\textwidth]{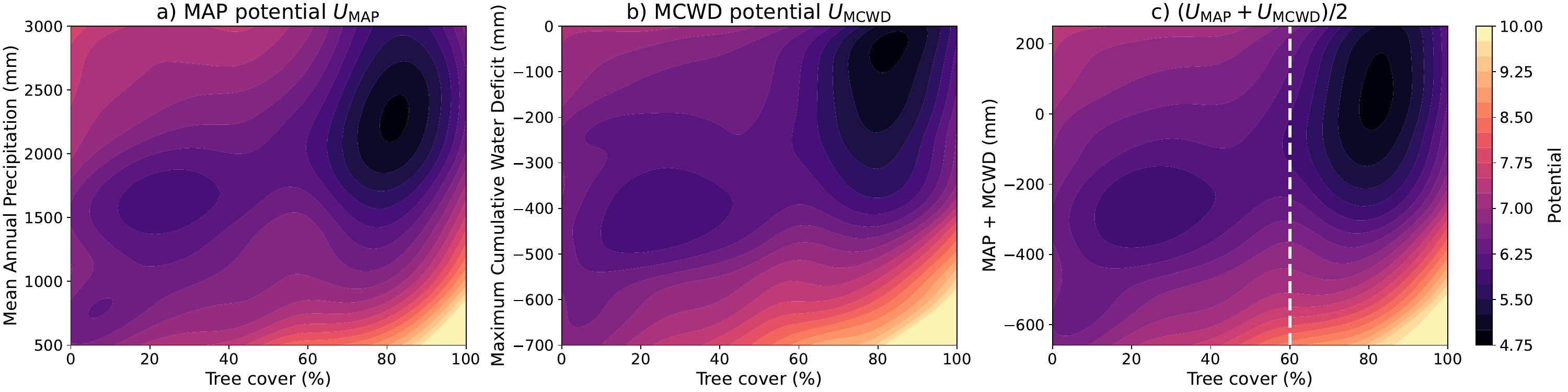}
	\caption{Potentials obtained from the data from~\cite{Flores2024} after application of Gaussian KDE, with the bandwidth detailed in Sec.~\ref{sec:potentials}. In all panels, a darker color indicates a smaller potential value. The left panel shows the potential corresponding to the Mean Annual Precipitation (MAP), while the middle panel shows the potential corresponding to the Maximum Cumulative Water Deficit (MCWD). The right panel presents the combined potential driving the Amazon model, where the $y$-axis of the Mean Annual Precipitation (MAP) has been rescaled from mm/year to mm/month to match that of MCWD. The white dashed line represents the target value used here as rainforest/low tree cover state threshold.}
	\label{fig:potentials}
\end{figure}

Following the method developed by~\cite{Livina2010} and already applied to the Amazon rainforest by~\cite{Hirota2011}, each distribution is transformed into a smoothed probability density function using Gaussian Kernel Density Estimation (KDE).
This method relies on the assumption that the tree cover is governed by an underlying stochastic differential equation:
\begin{equation}
	\mathrm{d}T = -\nabla U\mathrm{d}t + \epsilon\mathrm{d}W,
\end{equation}
where $U$ represents a potential, $W$ a Wiener process and $\epsilon$ its amplitude.
The corresponding Fokker-Planck equation then gives $U(T)=-\epsilon^2\log(p_e)/2$, where $p_e$ is the empirical probability density function.
\cite{Hirota2011} disregarded noise scaling to focus on the shape of the potential ($U/\epsilon^2$).
Here, the noise we introduce in the AMOC-Amazon model relates either to the variability of the AMOC (see Sect.~\ref{sec:amoc_model}) or to the fire process (see Sect.~\ref{sec:fire_term}), not directly to the tree cover dynamics, so we also focus on the sole shape of the potential.

However, the correct reconstruction of the probability density through KDE heavily depends on a so-called bandwidth parameter $h$, which determines the smoothness of the resulting distribution.
A too large value of $h$ may hide the distribution's structure by smoothing it too much, while a too small value results in the presence of data artifacts.
Like~\cite{Hirota2011}, we follow Silverman's rule of thumb: $h=1.06sn^{-1/5}$, where $n\sim3.10^5$ is the number of data points and $s$ is set to $3$, which corresponds to $0.03$ times the full range of tree cover (in percentage, between zero and $100$)~\cite{Hirota2011}.
We found empirically that, to preserve the bimodality of the $T$-MAP and $T$-MCWD distributions, $s$ had to be smaller than $6$.
Choosing $s=3$ provides a good trade-off between the smoothing effect of too large a bandwidth (that would eliminate the rainforest-low tree cover bistability) and the noisy effect of a too-small one (that would create spurious local minima).
For comparison, the equivalents of Fig.~\ref{fig:potentials}, but computed with $s=1,2,5$ and $7$ (respectively Fig.~\ref{fig:pot_1}, Fig.~\ref{fig:pot_2}, Fig.~\ref{fig:pot_5}, Fig.~\ref{fig:pot_7}) are shown in Appendix~\ref{app:potentials}.

The resulting potentials are shown in Fig.~\ref{fig:potentials}.
Panels a), b) and c) respectively present the MAP potential $U_{MAP}$, the MCWD potential $U_{MCWD}$ and the average of both.
All potentials have three local minima (indicated by a darker color): one around $80\%$ tree cover, corresponding to a rainforest state; another one around $25\%$ tree cover corresponding to a low tree cover state; and a third one at $0\%$, corresponding to a treeless state.
Here, however, we disregard the treeless state to focus on the transition from a rainforest to a low tree cover state.
\cite{Hirota2011} set the threshold between rainforest and low tree cover to $60\%$ mean tree cover.
In the rightmost panel of Fig.~\ref{fig:potentials}, we find that the $60\%$ threshold (represented by the white dashed line) roughly corresponds to the basin boundary between both local minima.
Therefore, we use this value as a target below: as soon as the mean tree cover in the Amazon rainforest has decreased down to $60\%$, we consider that it has transitioned to a low tree cover state, corresponding to a degraded forest.

\subsubsection{Wildfires as a noise process}
\label{sec:fire_term}

The third term in Eq.~\ref{eq:amazon_model} is a stochastic term.
It is meant to represent processes occurring on a faster time scale than the numerical time step, in particular the outburst of wildfires.
Since wildfires may occur abruptly and quickly destroy a large portion of the tree cover, \cite{Zheng2025} suggested modeling them as an $\alpha$-stable Lévy process, denoted as $\eta^\alpha$.
Such a noise process exhibits jumps, and its distribution is heavy-tailed, which is fit to describe the devastating impact within a short time span of fire outbursts.
An $\alpha$-stable Lévy process mainly depends on two parameters $\alpha$ and $\beta$.
The first one, $\alpha$, which belongs to the interval $(0,2]$, is called stability parameter and determines the ``width'' of the associated probability density function, while $\beta$, in the interval $[-1,1]$, determines the symmetry of the distribution.
Here, this noise process is meant to represent the destruction of tree cover due to wildfires.
Therefore, we cannot have positive noise values (fire cannot create trees), so we have to make sure that the noise distribution is one-sided towards negative values.
This is only possible when the parameters are set such that
$0<\alpha<1$ and $\beta=-1$.
We do not tune the value of $\alpha$: following~\cite{Zheng2025}, we take $\alpha=0.5$.
Moreover, we multiply $\eta^\alpha$ by a ``dampening'' parameter $\sigma$ to reduce the most extreme values of the $\alpha$-stable process.
We set $\sigma=0.5$.

\begin{figure}
	\centering
	\includegraphics[width=0.7\textwidth]{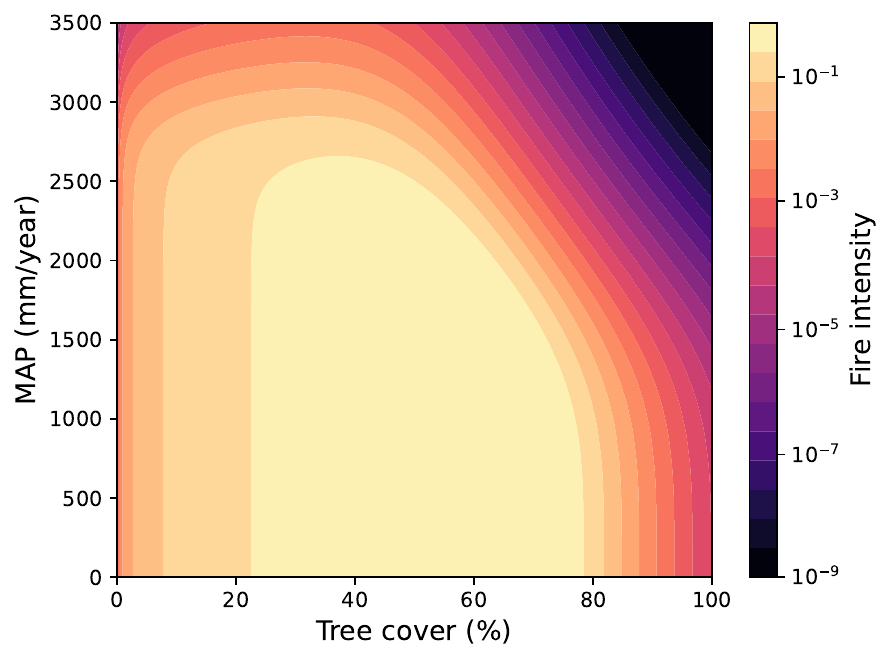}
	\caption{Fire intensity $f$ (Eq.~\ref{eq:fire_intensity}) as a function of tree cover and Mean Annual Precipitation (MAP).}
	\label{fig:fire_intensity}
\end{figure}

To represent fire damage, the scaled $\alpha$-stable process has to be multiplied by a physical function describing such damage.
The portion of tree cover destroyed by fires is described by $f(T,\mathrm{MAP})$~\cite{Staal2015}, as a function of tree cover ($T$) and precipitation ($\mathrm{MAP}$):
\begin{equation}
\label{eq:fire_intensity}
\begin{cases}
	f(T,\mathrm{MAP}) &= T\frac{I{(T,\mathrm{MAP})}^\gamma}{h_I^\gamma+I{(T,\mathrm{MAP})}^\gamma} \\
	I(T,\mathrm{MAP}) &= C(T)\times SMI(\mathrm{MAP}) \\
	C(T) &= \frac{h_C^\beta}{h_C^\beta + T^\beta} \\
	SMI(\mathrm{MAP}) &= \frac{h_{SMI}^\delta}{h_{SMI}^\delta + \mathrm{MAP}^\delta}
\end{cases}
\end{equation}
Fire damage $f$ is proportional to the tree cover $T$, multiplied by a function of the fire intensity $I$.
The fire intensity is represented by a product of two terms: landscape continuity $C$, which only depends on $T$ and soil moisture index $SMI$, which only depends on $\mathrm{MAP}$.
The main idea behind this model is that tree cover is left almost untouched by fire (i.e. $f$ is close to $0$) as long as the fire intensity is smaller than a certain threshold.
On the other hand, the forest is almost completely destroyed (i.e. $f$ is close to $T$) as long as the fire intensity is larger than a certain threshold.
To represent this behaviour, the fire damage $f$ uses a Hill function of the fire intensity $I$, of the form $I/(h_I+I)$, where $h_I$ represents a critical threshold for $I$.
Such function undergoes the largest change when $I$ is equal to the critical threshold $h_I$, and saturates towards $0$ (respectively, towards $1$) when $I$ becomes very small (respectively, very large).
The threshold $h_I$ therefore represents a fire intensity threshold for which the portion of destructed forest rapidly increases.
Both components of the fire intensity $I$ are also described by Hill functions, to represent the fact that they also undergo a large change when they cross a critical threshold.
However, the fire intensity increases as the tree cover $T$ and the precipitation $\mathrm{MAP}$ decrease.
Therefore, the Hill function used for $C(T)$ and $SMI(\mathrm{MAP})$ tends to one when its variable is very small (unlike the Hill function used for $f$).
The landscape continuity $C(T)$ accounts for the fact that fire is mainly fueled by grass because areas of open canopy make it easier for the fire to percolate through the forest.
There is thus a tree cover threshold $h_C$ beyond which the forest is too dense and the fire intensity drops.
The soil moisture index term $SMI(\mathrm{MAP})$ accounts for the effect of precipitation on the soil: the smaller the MAP, the dryer the soil and the grass, so the more intense the fire is.
The parameter $h_{SMI}$ represents the threshold beyond which the soil is wet enough and the fire intensity drops.
All parameters of the fire damage $f$ are presented in Table~\ref{tab:params} and this function is plotted in Fig.~\ref{fig:fire_intensity}.

Finally, \cite{Staal2015} indicate that fires cannot occur at any time.
Indeed, at a given location in the forest, a fire is not expected to outburst just a few days after the latest fire.
New fires can only appear after the forest has had time to regrow sufficient tree cover.
Therefore, \cite{Staal2015} introduced a ``fire return interval'' of seven years: on average, a given area of the forest undergoes a fire every seven years.
In practice, the fire return interval can be seen from another viewpoint.
Let $\mathrm{d}t$ be the numerical time step of the model, expressed in years.
For every latitude $\theta$ (we model the tree cover in the meridional direction only), there is at every time step a probability $\mathrm{d}t/7$ that a fire occurs.
So, aside from the $\alpha$-stable Lévy process used to model the intensity of fires, the occurrence of fires can also be modeled by a stochastic process.
We use a binomial process $\mathcal{B}(n,p)$ where $n$ is the total number of simulated time steps and $p=\mathrm{d}t/7$ is the probability of occurrence of a wildfire.

In practice, wildfires are modeled in the following way.
At each time step, we run on all grid points the binomial process determining the occurrence of a fire.
This process returns one on a certain number of grid points: it means that a fire occurs on all those grid points and at the current time step.
For each of the points where a fire occurs, we then generate a fire intensity from the $\alpha$-stable Lévy process $\eta^{\alpha}$.
This intensity is then scaled by $\sigma\times f(T,\mathrm{MAP})$ (Eq.~\ref{eq:fire_intensity}) to give it physical meaning.
Finally, we feed these fire intensities at the grid points where a fire occurs into the equation for tree cover (Eq.~\ref{eq:amazon_model}).

\subsection{Coupling of the AMOC and Amazon models through CESM}
\label{sec:cesm}

Until now, we have introduced the conceptual models of the AMOC and the Amazon rainforest.
In this Section, we describe how the coupling of the AMOC and the Amazon model is derived from CESM data.
The Amazon model is controlled by MAP and MCWD, but does not simulate these variables.
Instead, MAP, MCWD and their relation to the AMOC strength are computed in CESM for two region of the Amazon basin.
Using these relations, we can derive a value of MAP and MCWD for every AMOC strength simulated using the AMOC conceptual model.
These inferred time series of MAP and MCWD are then plugged into the Amazon model to simulate the tree cover in the selected regions.

\begin{figure}
	\centering
	\includegraphics[width=\textwidth]{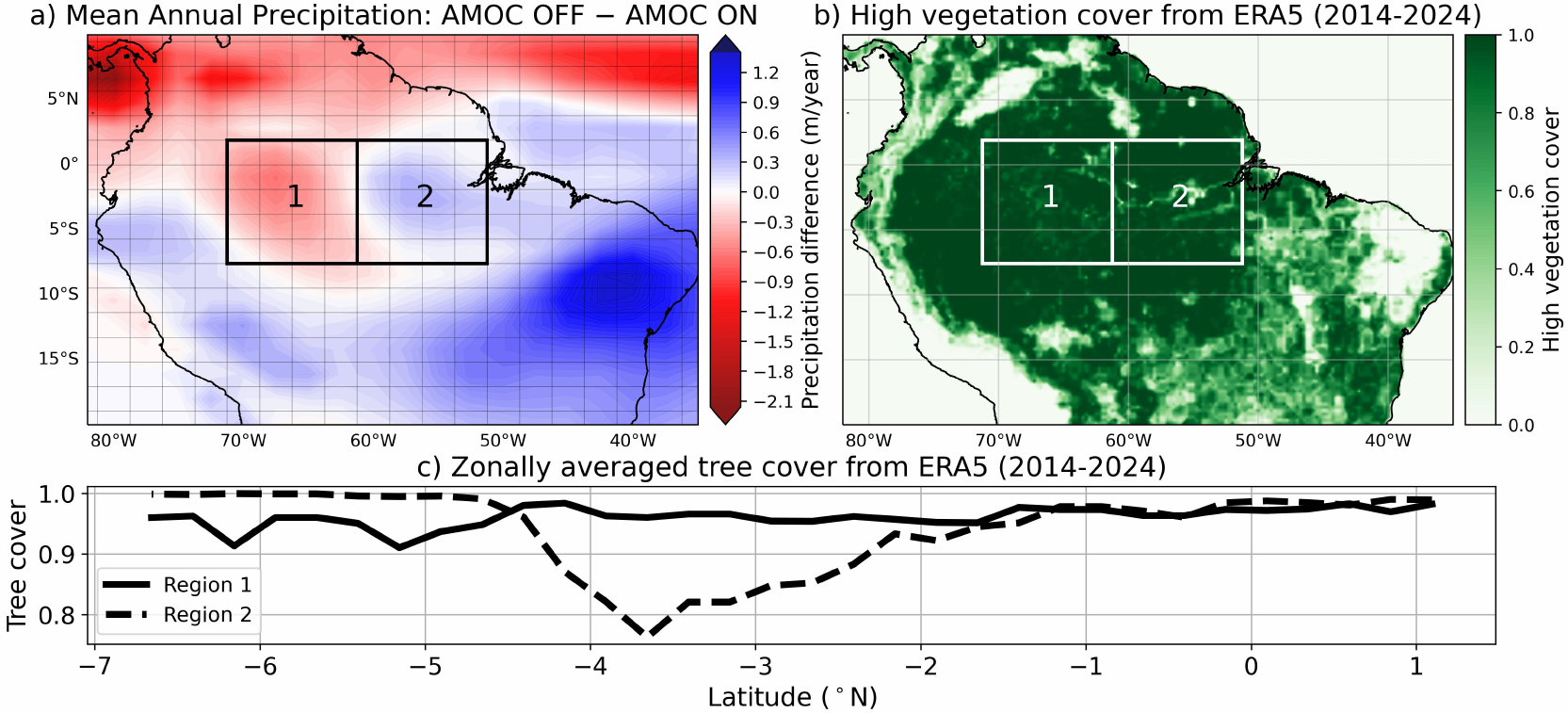}
	\caption{a) MAP difference in CESM over the Amazon between the AMOC collapsed state and the AMOC on-state. The grid represents CESM grid cells. Red areas indicate a decrease in precipitation due to the AMOC collapse. Blue areas indicate an increase in precipitation due to the AMOC collapse. The two studied regions are enclosed in solid lines and numbered.\\
	b) High vegetation cover from ERA5, averaged over the period $2014$-$2024$. The darker shades of green indicate a denser forest. The same regions as in panel a are enclosed in white lines.\\
	c) Zonally averaged tree cover from panel b in region 1 (solid line) and region 2 (dashed line). These curves are used as initial condition of the partial differential equation in Eq.~\ref{eq:amazon_model}.}
	\label{fig:regions}
\end{figure}

Figure~\ref{fig:regions}a presents the MAP difference over the Amazon in the collapsed AMOC state in CESM, compared to the AMOC on-state.
In this panel, the plotted grid is that of CESM.
When the AMOC collapses, the northwest of Brazil becomes dryer (shown in red in Fig.~\ref{fig:regions}a) while the Brazilian coastal region becomes wetter (shown in blue in Fig.~\ref{fig:regions}a).
Here, we focus on these two regions of the Amazon rainforest, highlighting the different impacts that an AMOC collapse may have on precipitation over the rainforest.
Figure~\ref{fig:regions}b presents the high vegetation cover from ERA5 over South America, averaged over the period 2014-2024.
The selected regions are enclosed in solid black lines in Fig.~\ref{fig:regions}a and in white lines (for readability) in Fig.~\ref{fig:regions}b and labeled $1$ and $2$.
As can be seen in Fig.~\ref{fig:regions}a, both regions span five grid cells of CESM in both meridional and zonal directions.
Region $2$ is the only region experiencing a wettening when the AMOC collapses while being part of the rainforest itself.
Figure~\ref{fig:regions}c presents the zonally averaged high vegetation cover in regions 1 (solid line) and 2 (dashed line).
These curves are used as initial condition for our conceptual Amazon model.

\begin{figure}[t]
	\centering
	\includegraphics[width=\textwidth]{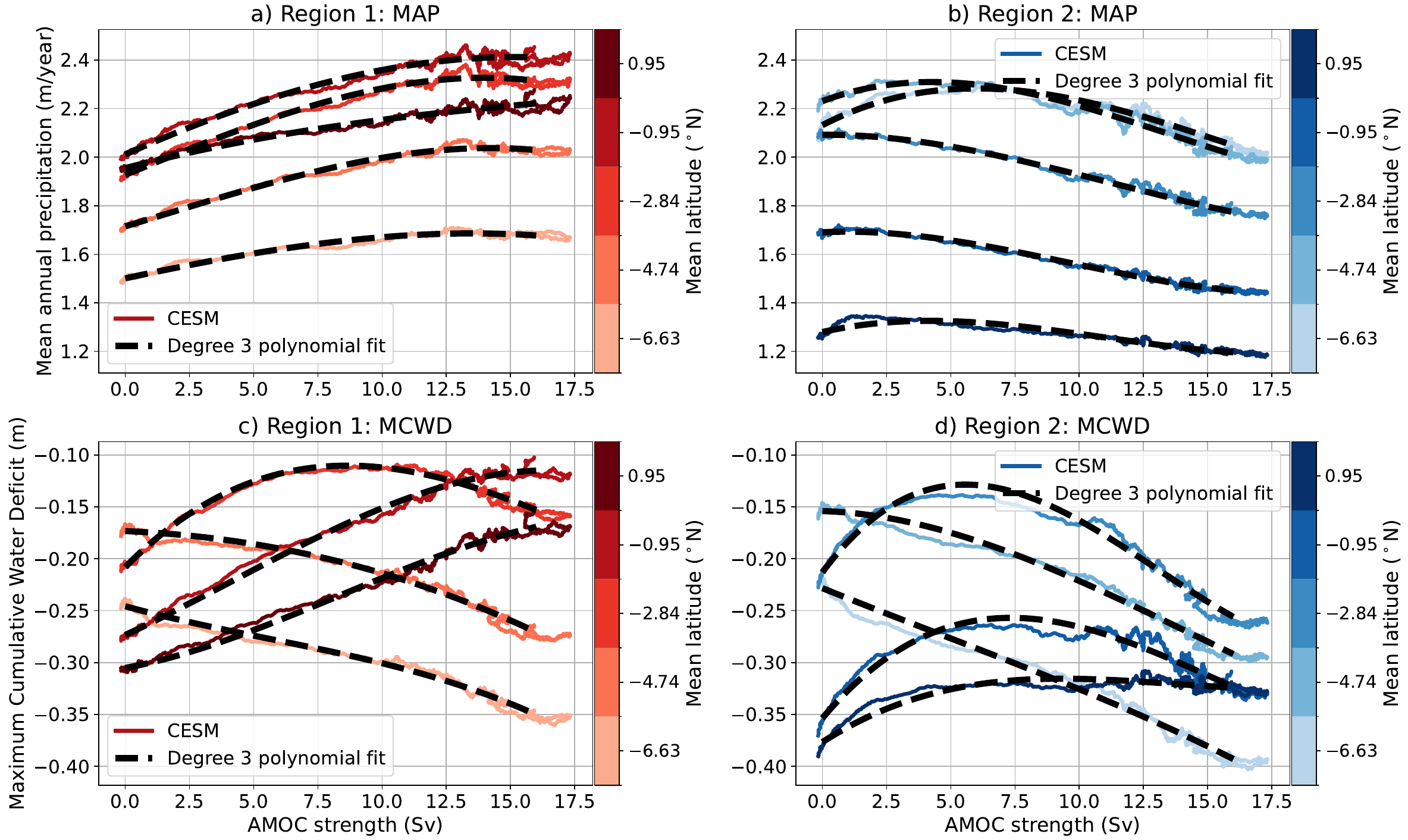}
	\caption{Zonally averaged MAP and MCWD from CESM, plotted against the AMOC strength for each grid cell of the two regions highlighted in Fig.~\ref{fig:regions}a. The upper row corresponds to MAP, the lower to MCWD. The left column corresponds to region 1, the right to region 2. Colored lines represent MAP and MCWD computed in CESM and smoothed by applying a $100$-year rolling mean, plotted against the AMOC strength from CESM, smoothed in the same way. The darker the line color, the larger the mean latitude of the corresponding grid cell in region 1 or 2 (see Fig.~\ref{fig:regions}a). Dashed lines indicate the least square fit of the corresponding curve by a degree three polynomial.}
	\label{fig:map_mcwd}
\end{figure}

To compute the values of MAP and MCWD over regions 1 and 2, precipitation in CESM is zonally averaged over these regions, in accordance with the conceptual AMOC model in Eq.~\ref{eq:amazon_model}.
In the meridional direction, regions 1 and 2 both span five grid cells of the CESM grid.
Therefore, we compute five values of MAP and MCWD, one for each grid cell.
Since MAP and MCWD are yearly quantities, they are plotted against the yearly averaged AMOC strength throughout the whole AMOC collapse simulated in CESM.
For each of the five grid cells, MAP and MCWD are plotted in Fig.~\ref{fig:map_mcwd} against the AMOC strength.
The upper row of Fig.~\ref{fig:map_mcwd} corresponds to MAP, the lower row to MCWD.
The left column of Fig.~\ref{fig:map_mcwd} corresponds to region 1, the right column to region 2.
In each panel, the colored lines present the evolution of MAP or MCWD against the AMOC strength in region 1 or 2.
The shade of the color of each line represents the mean latitude of the corresponding grid cell: the darker the curve, the larger the latitude (in $^\circ$N).
Because the raw output of CESM is very noisy, MAP, MCWD and the AMOC strength are smoothed by averaging them over a rolling window of $100$ years (i.e. $100$ data points).
We chose this window size so that the strength $\Psi_\mathrm{ON}$ of the AMOC on-state (defined as the average of the AMOC strength over the first $100$ years of simulation) corresponds to the first data point of the smoothed time series.

We have now obtained, for each of the five grid cells, an empirical relation between the hydrological variables and the AMOC strength.
But to determine in the coupled conceptual model which value of MAP and MCWD is associated with which AMOC strength, we need to express MAP and MCWD as a simple function of the AMOC strength.
To that end, we extrapolate the empirical relation by performing a least-square polynomial fit of the colored curves in Fig.~\ref{fig:map_mcwd}.
Each curve is independently fitted to a polynomial of degree three, which is the smallest degree allowing for a satisfying fit.
The dashed black lines in Fig.~\ref{fig:map_mcwd} present the resulting polynomials as a function of the AMOC strength over the interval $[0,\Psi_\mathrm{ON}]$.

In practice, a trajectory in the coupled AMOC-Amazon conceptual model is simulated as follows.
First, an AMOC trajectory is simulated using the conceptual model presented in Sect.~\ref{sec:amoc_model}.
Since no back coupling is implemented from the Amazon to the AMOC, the entire AMOC trajectory can be simulated in advance.
At every time step, we compute from the AMOC trajectory the corresponding values of MAP and MCWD for every grid cell, using the polynomial functions presented in Fig.~\ref{fig:map_mcwd}.
MAP and MCWD are then linearly interpolated between their five estimated values to obtain an array of MAP and MCWD compatible with the chosen resolution ($\mathrm{d}\theta=0.04^\circ$, see Sect.~\ref{sec:amazon_model}) to simulate the tree cover $T$ using the partial differential equation of Eq.~\ref{eq:amazon_model}.
Finally, the resulting arrays of MAP and MCWD are plugged into Eq.~\ref{eq:amazon_model}, which is initialized using the zonally averaged tree cover in region 1 or 2, as presented in Fig.~\ref{fig:regions}b.

We also determine for each region the equilibrium tree cover corresponding to the AMOC on-state, which will be useful to define a normalized tree cover (below, Sect.~\ref{sec:algo}).
The equilibrium cover is obtained by forcing Eq.~\ref{eq:amazon_model} with the MAP and MCWD corresponding to a constant AMOC strength $\Psi_\mathrm{ON}$ (obtained by averaging the AMOC strength over the first $100$ years of the CESM simulation), and without any fires.
The equilibrium tree cover in the AMOC on-state is equal to $75\%$ for region $1$ and to $73\%$ for region $2$.

\section{Trajectory Adaptive Multilevel Splitting (TAMS)}
\label{sec:tams}

Our main objective is to compute the cascading tipping probability of the coupled AMOC-Amazon system within a time horizon $t_{\max}$.
In other words, we attempt to compute the probability that the Amazon rainforest transitions to a low tree cover state after the AMOC has collapsed, within $t_{\max}$.
Since the AMOC and the Amazon conceptual models are stochastic, we focus on noise-induced transitions in both cases and do not vary any parameter that could trigger a bifurcation- or rate-induced tipping.
The probability of occurrence of these noise-induced transitions is estimated using Trajectory Adaptive Multilevel Splitting (TAMS).
The detailed algorithm is available in Appendix~\ref{app:tams} and we describe in the next subsection (Sect.~\ref{sec:algo}) its general principle only.

As explained by~\cite{Cerou2019a}, TAMS can estimate a large class of dynamical quantities, such as the mean transition time of the Amazon rainforest, and the distribution of the AMOC strength at every stage of the transition of the Amazon rainforest.
For that reason, when describing TAMS in the next Section (Sect.~\ref{sec:algo}), we present the estimation of a general function $\mathcal{O}(\mathbf{X})$ (where $\mathbf{X}$ designates a trajectory), rather than a probability.
In Sect.~\ref{sec:observables}, we then present the different forms that $\mathcal{O}$ takes for the present analysis of the impact of the AMOC on the Amazon rainforest.
Finally, we present in Sect.~\ref{sec:setup} the complete experimental setup used here to study the coupled AMOC-Amazon system.

\subsection{Description of the algorithm}
\label{sec:algo}

TAMS is a ``rare-event'' algorithm: it samples rare transitions much more efficiently than by direct Monte-Carlo simulation, by biasing ensemble simulations in a controlled way.
The key idea is that each trajectory $\mathbf{X}$ simulated during TAMS is assigned a weight $W_\mathbf{X}$~\cite{Brehier2016} such that the total sum of weights remains equal to one.
At every iteration of TAMS, the ensemble of trajectories is biased to favor larger decreases in tree cover.
Trajectories where the tree cover remains large are thus discarded while more and more unlikely trajectories (exhibiting a more abrupt decrease in tree cover) are sampled.
At every iteration, new trajectories are added to the total pool of trajectories (which also includes the discarded ones) in such a way that the sum of weights of the entire pool must remain one.
Since discarded trajectories cannot be modified anymore, their weight remains fixed and therefore the weights of all retained trajectories are decreased.
In this way, $W_\mathbf{X}$ effectively represents the likelihood of trajectory $\mathbf{X}$ and TAMS iteratively explores the tail of the distribution of trajectories.
TAMS can then estimate the expectation of any function $\mathcal{O}$ using the following formula~\cite{Brehier2016}:
\begin{equation}
	\label{eq:ams_formula}
	\hat{\mathcal{O}} = \sum_{\mathbf{X}} W_\mathbf{X}\mathcal{O}(\mathbf{X}),
\end{equation}
where the sum is taken over all trajectories $\mathbf{X}$ simulated during TAMS.
Equation~\ref{eq:ams_formula} is a weighted average of the function $\mathcal{O}$.
The average is taken over an approximation of the distribution of trajectories, where the likelihood of trajectory $\mathbf{X}$ (or the probability it is randomly simulated) is approximated by $W_\mathbf{X}$.
The function $\mathcal{O}$ can a priori represent any quantity, such as the averaged tree cover over a given area or time interval, the AMOC strength, or the probability that a wildfire occurs at a specific time step.
Since the sum of all weights equals one, Eq.~\ref{eq:ams_formula} simply performs a weighted average over all sampled values of $\mathcal{O}(\mathbf{X})$.
Each run of TAMS provides a different value of $\hat{\mathcal{O}}$ and it can be shown~\cite{Brehier2016} that their average is unbiased and converges to the true mean of $\mathcal{O}$.

To bias the ensemble simulation, TAMS relies on a score function acting as the distance to the transition, denoted here as $\varphi$.
In the case of the Amazon rainforest, there is a monotonous relation between a decrease in tree cover and a change in the system state, from rainforest to a degraded forest.
We can thus use the tree cover as score function.
For ease of use, the score function $\varphi$ is inverted and normalized so that it increases between $0$ and $1$ as the Amazon transitions from a rainforest to a degraded forest state.
Therefore, a normalized score $\varphi$ of zero corresponds to the equilibrium mean tree cover (see Sect.~\ref{sec:cesm}) of the Amazon model in the absence of fire and forced by a constant AMOC on-state (with a strength $\Psi_\mathrm{ON}$).
Note that trajectories are initialized at a larger tree cover than the equilibrium tree cover (corresponding to a score of zero), so, with our choice of setup, the initial condition has a negative score.
It does not make any difference for the algorithm, but it is more meaningful to assign a score of zero to an equilibrium tree cover rather than to an arbitrary initial condition.
The normalized score $\varphi$ is equal to one when the mean tree cover is equal to $60\%$ (see Sect.~\ref{sec:amazon_model}), which is the threshold defining a degraded forest state.
In other words, whenever the Amazon rainforest has a normalized score of $z$, it means that it has already lost $100\times z\%$ of the tree cover it has to lose to transition to a degraded forest.
In this view, the inverted and normalized score corresponds to a ``degradation function'': the degradation of the forest increases from $0$ to $1$ as the rainforest becomes a degraded forest.
In what follows, the score function $\varphi$ designates such degradation function, and any of its level $z$ is called ``degradation level''.
The goal of TAMS is to bring all trajectories in the ensemble simulation to strictly reach the target, which corresponds here to reaching a tree cover $T<0.6$, i.e. a degradation level strictly larger than one.

TAMS is initialized by simulating a set of $N$ independent trajectories until they reach a degraded forest state (i.e. a score of one) or until $t_{\max}$.
The score function is applied to each trajectory and computed at every time step.
We consider for each trajectory its maximum score: it indicates how close to the degraded forest state the system arrived before $t_{\max}$.
With our definition of the score function for the Amazon rainforest, the larger the score, the closer the forest to its degraded state.
Trajectories exhibiting the $n_c$ (which is a parameter of TAMS) lowest maximum scores are deemed the ``least successful'' ones in reaching the degraded state and are discarded.
But, since the numerical simulation of the system is discrete (due to the finite time step and spatial step and to the finite machine precision when evaluating the score), several trajectories may in specific cases have the same maximum score.
In practice, we denote as $\hat{n}_c$ the number of trajectories discarded at each iteration, which may be strictly larger than $n_c$.

Then, to keep an ensemble of fixed size, $\hat{n}_c$ new trajectories have to be simulated, replacing all discarded trajectories.
But we also have to ensure that these new trajectories are at least as ``successful'' as the trajectories that were retained.
In other words, we must ensure that the maximum score of the newly simulated trajectories is strictly larger than that of the discarded trajectories.
To do that, new trajectories are simulated by cloning retained trajectories until a branching point, after which the simulation is independently continued.
Let us call $\mu$ the largest of the $n_c$ smallest maximum scores.
So, the maximum score $\varphi_d$ of any discarded trajectory $\mathbf{X}_d$ is such that $\varphi_d\leq\mu$.
To replace $\mathbf{X}_d$ with a new trajectory, we randomly pick a trajectory $\mathbf{X}_c$ among the $N-\hat{n}_c$ remaining ones.
The maximum score $\varphi_c$ of $\mathbf{X}_c$ is strictly larger than $\mu$; otherwise $\varphi_c$ would have been part of the $n_c$ smallest maximum scores, and $\mathbf{X}_c$ would have been discarded.
Therefore, there exists at least one time step $s$ such that $\varphi(\mathbf{X}_c(s))>\mu$.
The trajectory $\mathbf{X}_c$ is cloned until the first time step $s$ (included) such that this condition is verified.
The cloned trajectory is then branched off $\mathbf{X}_c$ and simulated independently from time $s+1$ until time $t_{\max}$.
The cloned trajectory is however stopped early if it reaches a score of one (degraded forest state): in that case, a complete transition has been sampled.
Let $\mathbf{X}_n$ be the newly simulated trajectory.
Since $\mathbf{X}_n$ is cloned from $\mathbf{X}_c$ until time $s$ included, we know that $\varphi(\mathbf{X}_n(s))>\mu$.
Therefore, the maximum score of $\mathbf{X}_n$ is necessarily strictly larger than the maximum score of all discarded trajectories.
Moreover, since our coupled model is stochastic, $\mathbf{X}_n$ is independent from its parent $\mathbf{X}_c$ from time $s+1$ until the end of the trajectory.
All discarded trajectories are replaced by cloning a randomly chosen trajectory and branching it after their score strictly crosses $\mu$.
The weights of the discarded trajectories will never again be updated, assuming that their position in the distribution of trajectories has been found.
The weights of the ``surviving'' trajectories are updated and can be seen as their probability of survival until the current iteration.
The branched trajectories are given the updated weight of their parent.
Then, we compute the new $n_c$ smallest maximum scores and restart the cloning/resimulation procedure until obtaining $\mu=1$, where $1$ is by convention the target value of the score.

We call a ``run'' of TAMS the entire procedure, from the initialization, through every iteration, to termination of the algorithm.
For any function $\mathcal{O}$ of interest, each run of TAMS returns an estimate $\hat{\mathcal{O}}$ of its true expectation $\mathbb{E}[\mathcal{O}]$.
Therefore, we have to compute $M$ runs of TAMS to obtain a distribution of values of $\hat{\mathcal{O}}$.
\cite{Brehier2016} showed that the expectation $\mathbb{E}[\hat{\mathcal{O}}]$ of this distribution is equal to $\mathbb{E}[\mathcal{O}]$.

\subsection{Quantities of interest in the simulated ensembles}
\label{sec:observables}

\begin{figure}[t]
	\centering
	\includegraphics[width=\textwidth]{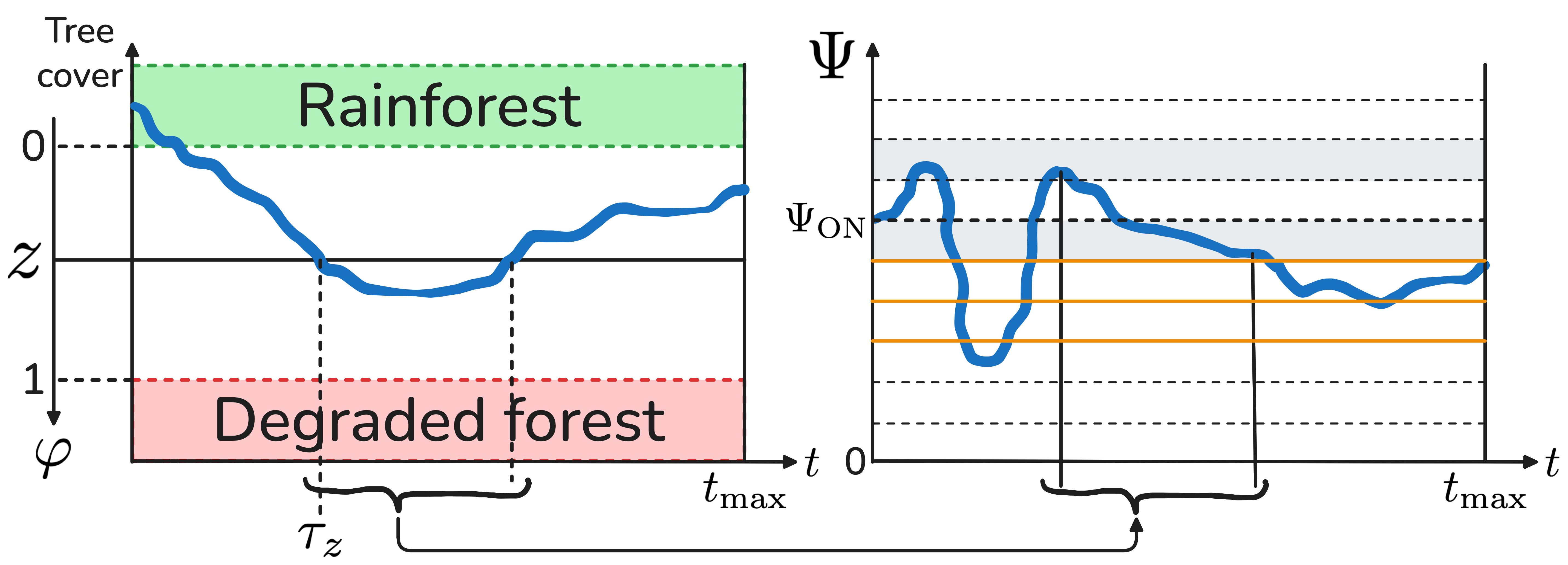}
	\caption{Summary of all quantities estimated using Eq.~\ref{eq:ams_formula} in the trajectories of the coupled conceptual model.
	The left panel presents a trajectory of the Amazon model (in terms of its mean tree cover), and the right panel shows the corresponding trajectory in the AMOC model (in terms of its AMOC strength $\Psi$).
	On the left panel, $\varphi$ is the degradation function, used as score function driving TAMS, and $z$ designates an arbitrary level of $\varphi$.
	The degradation function crosses $z$ at two time steps, shown by the horizontal dashed lines.
	Its first crossing defines the first-passage time $\tau_z$.
	Both crossings are taken into account to reconstruct the distribution of AMOC strengths when $\varphi$ crosses level $z$.
	These two time steps are therefore shown in the AMOC panel.
	On the right panel, $\Psi_\mathrm{ON}$ represents the AMOC on-state where the model is initialized.
	All horizontal dashed lines represent the bins of $\Psi$.
	The grayed bins represent the bins containing the values of $\Psi$ at the time steps where $\varphi$ crosses its level $z$.
	These bins result in a positive contribution to Eq.~\ref{eq:obs_distrib}.
	The orange horizontal lines show the values of $\Psi$ smaller than $\Psi_\mathrm{ON}$ (decreases in AMOC strength) that the AMOC strength crosses before $\varphi$ reaches its level $z$.
	These values result in a positive contribution to Eq.~\ref{eq:full_cond}.}
	\label{fig:summary}
\end{figure}

TAMS can be applied to any model where trajectories can be simulated and Eq.~\ref{eq:ams_formula} can be applied to any function of the model's trajectories.
Here, we apply TAMS to the AMOC-Amazon coupled conceptual model and use Eq.~\ref{eq:ams_formula} to estimate quantities that provide insight into the dynamics of the Amazon and into the AMOC impact on the Amazon.
These quantities are the probability $p$ that the Amazon transitions to a degraded forest before $t_{\max}$; the average time when this transition occurs; the evolution of the AMOC strength while the Amazon is transitioning; and the probability that the AMOC collapses before the Amazon rainforest transitions to a degraded forest.
In the following subsections, we detail the computation of each of these quantities from the trajectories of the coupled conceptual model, simulated through TAMS.
To illustrate these explanations, we will systematically refer to Fig.~\ref{fig:summary}.
Note that all of these quantities are estimated at every run of TAMS on the coupled model.
In particular when estimating the distribution of AMOC strength, we do not need to apply TAMS to the AMOC model separately: it is enough to run TAMS on the coupled model and count, using a well-defined function, the number of occurrences of the event of interest.

\subsubsection{Transition probabilities}
\label{sec:proba}

Here, we detail the estimation of the probability $p$ that the Amazon rainforest transitions to a degraded forest state, defined by $T<0.6$, before time $t_{\max}$.
Let $\mathbf{X}$ be a trajectory of the AMOC-Amazon coupled conceptual model.
We call $T(\mathbf{X})$ the mean tree cover at every time step.
Let $\tau_\mathrm{Amazon}$ be the first time step such that $T(\mathbf{X})<0.6$.
The time $\tau_\mathrm{Amazon}$ is termed ``transition time''.
With these notations, the transition probability $p$ can be reformulated as $p=\mathbb{P}(\tau_\mathrm{Amazon}<t_{\max})$.

In the last section, we have explained that TAMS can estimate the expectation of any function $\mathcal{O}$.
To do that, this function is applied to all trajectories $\mathbf{X}$ simulated through TAMS and we then compute the weighted average of the resulting values, using the weights $W_\mathbf{X}$ of the trajectories.
The computation of this weighted average corresponds to the application of Eq.~\ref{eq:ams_formula}.
Therefore, to estimate the transition probability $p$, we have to find a function $\mathcal{O}$ such that its expectation is equal to $p$:
$\mathbb{E}[\mathcal{O}]=\mathbb{P}(\tau_\mathrm{Amazon}<t_{\max})$.
This equality is satisfied by the function $\mathcal{O}(\mathbf{X})=\mathbbm{1}_{\tau_\mathrm{Amazon}<t_{\max}}(\mathbf{X})$, defined as:
\begin{equation}
	\mathbbm{1}_{\tau_\mathrm{Amazon}<t_{\max}}(\mathbf{X})=\begin{cases}
		1\ \mathrm{if}\ \tau_\mathrm{Amazon}(\mathbf{X})<t_{\max}\\
		0\ \mathrm{otherwise}
	\end{cases}.
\end{equation}
The function $\mathbbm{1}_\mathcal{C}(\mathbf{X})$, where $\mathcal{C}$ is a condition depending on trajectories $\mathbf{X}$, is called \textbf{indicator function} and will be used extensively to define quantities measured by TAMS.
In practice, for all trajectories simulated during the TAMS process, we save their weight and measure the corresponding value of $\mathcal{O}$.
By measuring the value of $\mathcal{O}$ on every trajectory simulated during TAMS, we can use Eq.~\ref{eq:ams_formula} to estimate the probability that the Amazon rainforest transitions to a degraded forest before $t_{\max}$.

However, the simple probability of the transition to a degraded forest does not give any information about the dynamics of the transition.
To obtain such insight into the transition itself, we can track the probability $p_z$ that the Amazon reaches every degradation level $z$ within $t_{\max}$.
For instance, let $z_1$ and $z_2$ two degradation levels such that $z_2>z_1$.
If we find that $p_{z_2}$ is much smaller than $p_{z_1}$, it means that reaching a degradation level $z_2$ is much more unlikely than reaching $z_1$.
Therefore, we can deduce that something has happened that prevented the forest from reaching $z_2$ after reaching $z_1$.
In this way, we obtain information about the dynamics of the transition at every degradation level, which we would not have with the single transition probability.

Moreover, estimating all values of $p_z$ is just as expensive as estimating $p$ only.
Whenever a trajectory reaches the degraded forest state ($z=1$), starting from the rainforest state ($z=0$), it must have crossed every intermediate level $z$ in the process.
Therefore, when simulating trajectories through TAMS to estimate the transition probability $p$, we also generate all the necessary information to estimate any probability $p_z$, with $z<1$.
There is a minimal increase in the memory cost of TAMS because we have to store as many scalars as the number of levels $z$ considered.
However, this extra cost is negligible compared to the cost of simulating trajectories.

To estimate the value of $p_z$ for a given level $z$, we introduce the first-passage time across a given level $z$ as
\begin{equation}
	\tau_z=\min\{t\in\mathbb{R}_+|\varphi(\mathbf{X}(t))>z\}.
\end{equation}
For instance, the trajectory shown in the left panel Fig.~\ref{fig:summary} crosses the degradation level $z$ before $t_{\max}$ and the corresponding first-passage time $\tau_z$ is indicated on the time axis.
If we take $z=1$, then the first-passage time $\tau_1$ is equivalent to $\tau_\mathrm{Amazon}$ as defined above.
Therefore, by analogy with the above discussion, we can estimate the transition probability $p_z$ to cross level $z$ before $t_{\max}$ by applying Eq.~\ref{eq:ams_formula} to the function $\mathcal{O}_z(\mathbf{X})=\mathbbm{1}_{\tau_z<t_{\max}}(\mathbf{X})$.
This function simply measures whether trajectory $\mathbf{X}$ has reached level $z$ before $t_{\max}$.
Note that we do not need to measure $\tau_z$ precisely: evaluating at a time $t$ that $\varphi(\mathbf{X}(t))>z$ implies that $\tau_z$ is well defined and smaller than $t$.
For instance, on the left panel of Fig.~\ref{fig:summary}, we find that the example trajectory, say $\mathbf{X}_\mathrm{ex}$, has crossed level $z$ before $t_{\max}$, so $\mathcal{O}_z(\mathbf{X}_\mathrm{ex})=1$.
Then, Eq.~\ref{eq:ams_formula} is applied to every function $\mathcal{O}_z$ to compute the probability of reaching any level $z$ of $\varphi$ before $t_{\max}$ :
\begin{equation}
\label{eq:pz}
p_z = \mathbb{P}(\tau_z<t_{\max})=\mathbb{E}[\mathcal{O}_z]=\mathbb{E}[\mathbbm{1}_{\tau_z<t_{\max}}].
\end{equation}

\subsubsection{Mean first-passage times (MFPT)}
\label{sec:mfpt}

We have just explained how to track the probabilities $\mathbb{P}(\tau_z<t_{\max})$ of crossing any level $z$ of the degradation function $\varphi$ before $t_{\max}$.
Another interesting quantity to analyze is the time scale of the transition of the Amazon to a degraded forest.
It amounts to estimating the expectation of the first-passage times $\tau_z$, as defined in the former subsection (Sec.~\ref{sec:proba}).
Their expectation is called Mean First-Passage Time (MFPT).
Indeed, the MFPT is a signature of the mechanism that explains the transition: we can expect the transition to unfold on different time scales if it is purely driven by an AMOC collapse or by intense wildfires.
Moreover, different mechanisms may be dominant at different stages of the transition.
To capture the full history of the transition, we can therefore estimate the time at which the Amazon reaches every level of the degradation function $\varphi$.

Here, we aim to estimate the expectation of $\tau_z$ across the ensemble simulation obtained with TAMS.
As was done for the probabilities $\mathbb{P}(\tau_z<t_{\max})$, we need to find a function $\mathcal{O}$ to which we can apply the weighted average defined in Eq.~\ref{eq:ams_formula}.
For a given trajectory $\mathbf{X}$, the function $\mathcal{O}$ has to capture $\tau_z(\mathbf{X})$ but only if the degradation level $z$ was reached in the trajectory $\mathbf{X}$.
Therefore, we actually want to compute the \textbf{conditional} expectation of $\tau_z$, conditioned on reaching the degradation level before $t_{\max}$.
Similar to the former subsection, we consider the indicator function $\mathbbm{1}_{\tau_z<t_{\max}}(\mathbf{X})$, which returns $1$ if trajectory $\mathbf{X}$ reaches the degradation level $z$ before $t_{\max}$ and $0$ otherwise.
The goal here is thus to estimate the expectation: $\mathbb{E}[\tau_z\,|\,\mathbbm{1}_{\tau_z<t_{\max}}=1]$.

Using the definition of a regular conditional probability, we can write:
\begin{equation}
\label{eq:mfpt}
	\mathbb{E}[\tau_z|\mathbbm{1}_{\tau_z<t_{\max}}]=\frac{\mathbb{E}[\tau_z\mathbbm{1}_{\tau_z<t_{\max}}]}{\mathbb{E}[\mathbbm{1}_{\tau_z<t_{\max}}]}=\frac{\mathbb{E}[\tau_z\mathbbm{1}_{\tau_z<t_{\max}}]}{p_z},
\end{equation}
where $p_z$ is the probability that the Amazon reaches a degradation level of $z$ before $t_{\max}$ (see Eq.~\ref{eq:pz}).
By applying the procedure described in the former subsection, we already have an estimate of $p_z$.
All that remains to estimate is thus the numerator of Eq.~\ref{eq:mfpt}.
This numerator is simply the expectation of a function $\mathcal{O}(\mathbf{X})=\tau_z(\mathbf{X})\mathbbm{1}_{\tau_z<t_{\max}}(\mathbf{X})$.
We can estimate this expectation by computing this function for all trajectories simulated by TAMS and applying the weighted average in Eq.~\ref{eq:ams_formula}.

Note that TAMS stops as soon as all trajectories in the ensemble have transitioned to a degraded forest before $t_{\max}$.
Therefore, all trajectories must go through all degradation levels $z$ before $t_{\max}$.
It is thus just as costly to run TAMS with or without estimating the MFPT $\tau_z$: the trajectories have to be simulated anyway across all levels $z$.
So, by estimating all MFPT, we make better use of the available data.

In the former subsection, the probabilities $p_z$ to cross every degradation level before $t_{\max}$ were estimated by computing the expectation of a certain function.
Here, however, we are interested in a conditional expectation, thus a ratio of expectations (see Eq.~\ref{eq:mfpt}).
Such conditional expectation is called a \textit{normalized} quantity~\cite{Cerou2019a}.
It is a crucial point, since Eq.~\ref{eq:ams_formula} is only unbiased when applied to non-normalized measures~\cite{Brehier2016,Cerou2019b} (such as the estimate of transition probabilities).
More specifically, \cite{Cerou2019b} indicate that the estimation of the normalized quantities using Eq.~\ref{eq:ams_formula} is biased with a factor $1/N$, where $N$ designates the size of the simulated ensemble.
Moreover, \cite{Cerou2019a} showed that both the estimates of non-normalized and normalized measures follow a central limit theorem in the limit of infinite $N$.
So, for a large enough value of $N$, the bias on the estimation of normalized quantities using Eq.~\ref{eq:ams_formula} is negligible.
Here, we use $N=1000$ (see Sect.~\ref{sec:setup}), which means that the bias on $\tau_z$ is much smaller than typical values of $\tau_z$ (of the order of several years).

\subsubsection{AMOC strength distribution}
\label{sec:distrib}

Until now, we have only considered the dynamics of a transition of the Amazon rainforest.
We will now look into the dynamics of the AMOC while the rainforest transitions, to gain later insight into what the impact of the AMOC on the Amazon might be.
The easiest way to analyze the dynamics of the AMOC is to directly look in the ensemble of trajectories at the distribution of AMOC strengths at every stage of the Amazon transition.
Indeed, such distribution may highlight the fact that certain AMOC strengths are preferred at certain stages of the Amazon transition.
It would then suggest that the AMOC strength must take certain values so that the Amazon can reach this stage of its transition.
Importantly, we do not include the AMOC strength in the score function driving TAMS because we do not want to bias simulations towards any specific AMOC dynamics.
Instead, we want to generate Amazon transitions and then find out what kind of AMOC behavior favored these transitions by analyzing the distribution of AMOC strength in the trajectories selected by TAMS.

The goal here is to reconstruct the distribution of AMOC strengths when the Amazon reaches a given degradation level $z$.
To do so, the only available information is the values of $\Psi$ that the AMOC takes at the time $\tau_z$ (when the level $z$ is crossed).
Of course, this reconstruction should only use data from trajectories that reach the level $z$ before $t_{\max}$.
The main idea is to chose a range $[\Psi_{\min},\Psi_{\max}]$ of values of $\Psi$ and divide it into $B$ bins of equal size, denoted as $r_1,\ldots,r_B$.
More precisely, the bin $r_i$ is defined by
\begin{equation}
r_i=[\Psi_{\min}+\frac{i-1}{B}(\Psi_{\max}-\Psi_{\min}),\Psi_{\min}+\frac{i}{B}(\Psi_{\max}-\Psi_{\min})].
\end{equation}
For instance, consider the trajectory shown in Fig.~\ref{fig:summary}.
Its AMOC component is presented in the right panel.
In this panel, the distance between each horizontal line represents a bin of the AMOC strength $\Psi$.

As was already done in the former subsections, we now have to define a function $\mathcal{O}$ such that its expectation gives the probability that $\Psi$ belongs to bin $r_i$ as the degradation function $\varphi$ crosses its level $z$.
Actually, there has to be one such function per bin $r_i$ and per level $z$, thus we denote these functions as $\mathcal{O}_{z,r_i}$.
Moreover, since a given trajectory $\mathbf{X}$ may cross level $z$ at any time step, and even cross it several times, the functions $\mathcal{O}_{z,r_i}$ have to consider all time steps of $\mathbf{X}$.
Finally, at every time step $t$, a state $\mathbf{X}(t)$ makes a non-zero contribution to $\mathcal{O}_{z,r_i}$ (and therefore to the distribution of $\Psi$) only if $\Psi(\mathbf{X}(t))$ belongs to the bin $r_i$ and if $\varphi(\mathbf{X})$ crosses its level $z$ around the time step $t$.
Let us look, for example, at the trajectory $\mathbf{X}_\mathrm{ex}$ in Fig.~\ref{fig:summary}.
In the left panel, its degradation function $\varphi$ crosses level $z$ at two time steps, indicated by the vertical dashed lines.
In the right panel, the bins $\Psi$ belongs to when $\varphi$ crosses its level $z$ are shown in gray, and called $r_i$ and $r_j$.
So, $\mathcal{O}_{z,r_i}(\mathbf{X}_\mathrm{ex})=1$ and $\mathcal{O}_{z,r_j}(\mathbf{X}_\mathrm{ex})=1$.
For all the other bins $r$, the functions $\mathcal{O}_{z,r}(\mathbf{X}_\mathrm{ex})$ are equal to zero.

To satisfy these constraints, we end up with the following expression for the functions $\mathcal{O}_{z,r_i}$:
\begin{equation}
	\label{eq:obs_distrib}
	\mathcal{O}_{z,r_i}(\mathbf{X}) = \sum_{t=0}^{\min\{\tau_\mathrm{Amazon}(\mathbf{X}),t_{\max}\}} \mathbbm{1}_{(\varphi(\mathbf{X}(t))-z)(\varphi(\mathbf{X}(t+\mathrm{d}t))-z)<0}(\mathbf{X})\mathbbm{1}_{\Psi(\mathbf{X}(t))\in r_i}(\mathbf{X}).
\end{equation}
In this formula, the sum is meant to take into account all time steps until the trajectory $\mathbf{X}$ stops, which only happens when reaching a degraded forest state (at time $\tau_\mathrm{Amazon}$) or at the maximum time $t_{\max}$.
At every time step, we evaluate a product of two indicator functions.
First, the function $\mathbbm{1}_{(\varphi(\mathbf{X}(t))-z)(\varphi(\mathbf{X}(t+\mathrm{d}t))-z)<0}(\mathbf{X})$ tests whether the degradation function $\varphi$ has crossed level $z$ at the time step $t$.
Second, the function $\mathbbm{1}_{\Psi(\mathbf{X}(t))\in r_i}(\mathbf{X})$ tests whether, at this time step, the AMOC strength $\Psi$ belongs to the bin $r_i$.
If both indicator functions return one, then one is added to the probability distribution estimate of bin $r_i$ across level $z$.

The expectation of the distribution of AMOC strengths at every stage of the Amazon transition is then obtained by applying Eq.~\ref{eq:ams_formula} to all functions $\mathcal{O}_{z,r_i}$ (as was done for the transition probabilities and MFPT).
However, we aim to estimate the distribution of $\Psi$ \textbf{conditioned} on crossing level $z$ (as was the case for the MFPT), which is a normalized quantity.
So, after TAMS has terminated, we have to divide all estimates $\mathbb{E}[\mathcal{O}_{z,r_i}]$ by $p_z$.
As for the MFPT, the estimated conditional expectation is biased, but the bias vanishes for $N$ sufficiently large.
Finally, when TAMS has terminated, we can retrieve a proper probability distribution by normalizing each conditional expectation $\mathbb{E}[\mathcal{O}_{z,r_i}]/p_z$ by $\sum_{i=1}^{B}\mathbb{E}[\mathcal{O}_{z,r_i}] / p_z$.

\subsubsection{Cascading tipping probabilities}

We now aim to estimate the probability that an AMOC collapse initiates a tipping cascade by triggering, or favoring, a transition of the Amazon rainforest to a degraded forest.
Note that, by design, TAMS only samples trajectories that reach a degraded forest state or stop at the time $t_{\max}$.
Therefore, the estimated cascading probability is necessarily conditioned on Amazon and AMOC transitions occurring before $t_{\max}$, and is thus dependent on $t_{\max}$.
To evaluate the impact of the AMOC on the Amazon, this cascading probability should compare the likelihood of any transition of the Amazon within $t_{\max}$, to the likelihood of that same transition when an AMOC collapse has already occurred.
We can thus think of the cascading probability as a conditional probability: the probability that the Amazon transitions \textbf{after} the AMOC does, given that the Amazon transitions within $t_{\max}$.

To define this conditional probability, we first need to define the time at which the AMOC collapses.
It is denoted as $\tau_\mathrm{AMOC}$ and corresponds to the first time such that the AMOC strength $\Psi$ equals zero.
Following the definition of a conditional probability, we then define the cascading probability as:
\begin{equation}
\label{eq:cond_proba}
	\mathbb{P}(\tau_\mathrm{AMOC}<\tau_\mathrm{Amazon}\ |\ \tau_\mathrm{Amazon}<t_{\max}) = \frac{\mathbb{P}(\tau_\mathrm{AMOC}<\tau_\mathrm{Amazon}<t_{\max})}{\mathbb{P}(\tau_\mathrm{Amazon}<t_{\max})}.
\end{equation}
This quantity counts the number of times that the AMOC collapses before the Amazon rainforest transitions, given that the latter transitions before $t_{\max}$.
We now detail its possible interpretations.
When it is close to one, either the AMOC is very likely to collapse on a faster time scale than the Amazon rainforest, or the AMOC collapse is a necessary condition for the Amazon rainforest to transition.
In the first case, the fact that this indicator is close to one is solely due to a difference in time scales between a fast AMOC and a slow rainforest, so not much can be deduced about the connection between both systems.
In that case, no causal link can be deduced from an AMOC collapse to an Amazon transition, so transition time scales have to be carefully assessed.
But, on the other hand, if the AMOC collapse is unlikely and/or slower than the Amazon transition, finding the probability in Eq.~\ref{eq:cond_proba} close to one across independent runs of TAMS means that all observed transitions of the Amazon are preceded by an AMOC collapse, although this order of events should be unlikely.
In this case, a probability of one is no proof of the existence of a tipping cascade, but it suggests the presence of a causal link between the AMOC collapse and the Amazon transition.

If the transitions of the two systems are not causally linked and both transitions occur on similar time scales, the  conditional probability in Eq.~\ref{eq:cond_proba} is expected to tend to $0.5$.
Indeed, the AMOC collapse may or may not occur before the rainforest transitions to a degraded forest.
Both transitions may be correlated (e.g. both caused by the collapse of another system), but one does not a priori cause the other.
Finally, a conditional probability close to zero indicates that either the AMOC cannot collapse (or collapses on a slower time scale than the Amazon transitions), or the Amazon transition cannot be preceded by an AMOC collapse.
The first possibility indicates a simple correlation.
But, if the AMOC is known to collapse on a time scale comparable to that of the Amazon transition, a conditional probability close to zero then indicates that an AMOC collapse hinders the transition of the Amazon, thus a negative causal link between both systems.

In all cases, the conditional probability in Eq.~\ref{eq:cond_proba} alone cannot prove or disprove the presence of a tipping cascade, and the time scales of the transitions of both the AMOC and the Amazon have to be carefully considered.
Therefore, this indicator is meaningful when accompanied by an analysis of both systems, which can be carried out using other functions estimated with TAMS, as described in Sec.~\ref{sec:proba}, \ref{sec:mfpt} and \ref{sec:distrib}.

However, the estimate of the joint probability (the numerator of Eq.~\ref{eq:cond_proba}) has a limited resolution.
Indeed, if we compute $K$ independent runs of TAMS each containing $N$ ensemble members, the minimum number of times we can observe the joint event is once over the ensemble of $KN$ trajectories exhibiting an Amazon transition.
Hence, the smallest joint probability we can estimate is $1/(KN)$.

However, as the AMOC weakens and the Amazon transitions, the relationship between both systems can evolve.
Moreover, if both transitions unfold over very different time scales, it is possible that one is very unlikely to transition before the other because it does not have the time.
As already mentioned, this situation can affect the interpretation of the cascading probability explained above.
One way to circumvent this issue is to obtain more insight into the history of the relation between both systems by estimating intermediate cascading probabilities all along the transition of the Amazon.
Namely, we can compare the probability that the Amazon undergoes any degradation within $t_{\max}$ to the probability that any AMOC weakening precedes this degradation.
In this way, we can track how the dynamics of the AMOC and the Amazon compare all along the Amazon transition.
Let us look for instance at the trajectory $\mathbf{X}_\mathrm{ex}$ in Fig.~\ref{fig:summary}.
Here, we are interested in measuring \textbf{decreases} in AMOC strength (compared to its strength $\Psi_\mathrm{ON}$ in the on-state) occurring before increases in the Amazon degradation $\varphi$.
On the right panel of Fig.~\ref{fig:summary}, the orange horizontal lines represent the discretized values of AMOC strength smaller than $\Psi_\mathrm{ON}$ that the system crosses before $\varphi$ first reaches level $z$.
Therefore, there may be a non-negligible probability that a strong AMOC weakening precedes a degradation of $z$.
In this case, such AMOC weakening may be part of the mechanism explaining that the Amazon reaches a degradation of $z$.
These intermediate conditional probabilities thus may enlighten the full history of the interaction between the AMOC and the Amazon forest.

It is clear that such intermediate cascading probability is a generalization of the conditional probability described in Eq.~\ref{eq:cond_proba}, where the complete transition of both systems is replaced with the crossing of intermediate levels.
To define this new cascading probability, let $z$ be a level of the degradation function $\varphi$ and $\psi$ be a value of the AMOC strength.
Consistent with $\tau_z$, we define the first time that the AMOC strength decreases down to $\psi$ as $\tau_\psi(\mathbf{X}) = \min\{t\in[0,t_{\max}]\ |\ \Psi(\mathbf{X}(t))<\psi\}$.
We now estimate the probability that any decrease in AMOC strength occurs before any increase in degradation, given that the latter occurs within $t_{\max}$:
\begin{equation}
	\label{eq:full_cond}
	\mathbb{P}(\tau_\psi<\tau_z\ |\ \tau_z<t_{\max}) = \frac{\mathbb{P}(\tau_\psi<\tau_z<t_{\max})}{\mathbb{P}(\tau_z<t_{\max})} = \frac{\mathbb{P}(\tau_\psi<\tau_z<t_{\max})}{p_z}.
\end{equation}
We will estimate this conditional probability for a range of values of $z$ in the interval $[0,1]$ (across the full Amazon transition) and a range of values of $\psi$.
First, note that the denominator in Eq.~\ref{eq:full_cond} is already known: it is exactly the transition probability of the Amazon rainforest to a degraded forest state (discussed in Sect.~\ref{sec:proba}).
So, to estimate the conditional probability in Eq.~\ref{eq:full_cond}, we only need to focus on its numerator.

To estimate the probability $\mathbb{P}(\tau_\psi<\tau_z<t_{\max})$, as was done before (see Sect.~\ref{sec:proba}), we have to find a function $\mathcal{O}$ such that $\mathbb{E}[\mathcal{O}]=\mathbb{P}(\tau_\psi<\tau_z<t_{\max})$.
By analogy with Sect.~\ref{sec:proba}, the function $\mathcal{O}$ should be one only when the Amazon reaches a degradation of $z$ before $t_{\max}$ \textbf{and} when the AMOC strength has already reached a value $\psi$.
Inspired by the estimator of $p_z$, we can design the following function:
\begin{equation}
	\label{eq:obs_cond_proba}
	\mathcal{O}_{z,\psi}(\mathbf{X}) = \mathbbm{1}_{\tau_z<t_{\max}}(\mathbf{X})\times\mathbbm{1}_{\tau_\psi<\tau_z}(\mathbf{X}).
\end{equation}
This function is a product of two indicator functions, and will thus be equal to one only if both are equal to one.
The first indicator function returns one if and only if the degradation function of the Amazon in trajectory $\mathbf{X}$ crosses its level $z$ before $t_{\max}$.
The second indicator function returns one if and only if the AMOC strength decreases down to a value $\psi$ before the degradation function crosses its level $z$.
Let us go back to the illustration in Fig.~\ref{fig:summary}.
On the right panel, we have $\mathcal{O}_{z,\psi}(\mathbf{X}_\mathrm{ex})=1$ for all levels $\psi$ of the AMOC strength corresponding to the orange lines, and we have $\mathcal{O}_{z,\cdot}(\mathbf{X}_\mathrm{ex})=0$ for all other levels of the AMOC strength.

\subsection{Experimental setup}
\label{sec:setup}

TAMS is applied to the AMOC-Amazon coupled conceptual model for both regions $1$ and $2$ (shown in Fig.~\ref{fig:regions}) using $N=1000$ trajectories.
At every iteration of the algorithm, we discard all trajectories whose maximum score is among the $n_c=10$ smallest maximum scores.
For each region, TAMS is run $K=20$ times independently.
We estimate from these runs $20$ independent expectations of the quantities introduced in Sect.~\ref{sec:observables}.
The final estimate of these quantities is the average of these expectations, presented with their $95\%$ confidence interval over these $20$ runs of TAMS.
As explained in Sect.~\ref{sec:algo}, TAMS is driven by the degradation function $\varphi$ as score function.
TAMS is applied to the whole coupled model, although $\varphi$ only depends on the tree cover.
Therefore, the AMOC follows its natural variability and is not directly subject to the selection pressure induced by TAMS on the mean tree cover.

The coupled model is simulated for $t_{\max}=200$ years with a time step $dt=0.05$ years.
The time horizon of $200$ years was chosen to leave enough time for the AMOC and the Amazon rainforest to transition, without making these events too likely.
If the time horizon is too short, the occurrence of the AMOC or Amazon transition would rely on extreme and very rare events.
Conversely, the larger the time horizon, the larger the transition probability of the AMOC and the Amazon because they are both forced by unbounded noise.
Therefore, the choice of $t_{\max}$ impacts all results presented below: for instance, if $t_{\max}$ was increased, all probabilities and mean first-passage times would likely increase.
In all sections below, the term ``transition'' only refers to reaching the degraded forest state (degradation function $z=1$), starting from the rainforest state (degradation function $z=0$).
But a transition is a dynamic process, so we can estimate all quantities introduced in Sect.~\ref{sec:observables} ``at every stage'' of the transition, meaning that they are estimated at any level of the degradation function.

\section{Results}
\label{sec:results}

When applying TAMS to the AMOC-Amazon coupled conceptual model, we can estimate the expectation of all quantities introduced in Sect.~\ref{sec:observables}.
The estimates of all quantities reconstructed with TAMS from trajectories of the coupled model are analyzed in the following sections, in the same order as they were presented in Sect.~\ref{sec:observables}.
We explain how they provide insights into the AMOC-Amazon dynamics and analyze the impact of the AMOC on the Amazon that can be deduced from the outputs of TAMS.

\subsection{Mean Amazon decline probabilities and mean first-passage times}

\begin{figure}
	\centering
	\includegraphics[width=\textwidth]{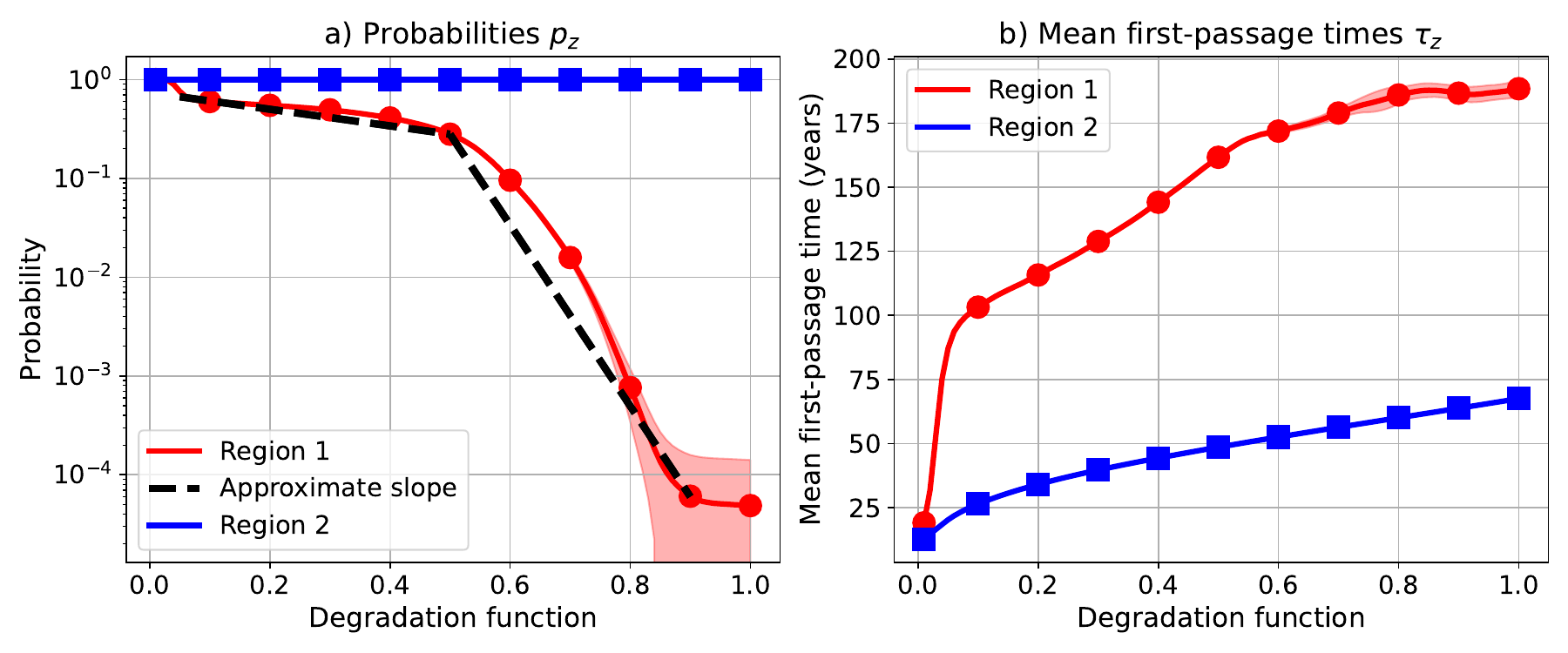}
	\caption{a) Probabilities $p_z=\mathbb{P}(\tau_z<t_{\max})$ (see Sect.~\ref{sec:proba}) that the degradation function of the Amazon reaches any level $z$ within $200$ years, for both regions shown in Fig.~\ref{fig:regions}. For $z=1$, it is the probability that the rainforest transitions to a degraded forest within $200$ years. Dashed black lines show the approximate slopes of $p_z$ in the different stages of the complete transition.\\
	b) Mean first-passage times $\tau_z$ (see Sect.~\ref{sec:mfpt}) across any level $z$ of the degradation function.\\
	In both panels, the solid lines are the expectations over $20$ independent TAMS runs. Markers indicate the tenths of the degradation function. The shaded areas represent the $95\%$ confidence intervals computed over the same runs.}
	\label{fig:proba_mfpt}
\end{figure}

We start by analyzing the probabilities that the Amazon rainforest reaches any degradation level $z$ before $t_{\max}$, denoted as $p_z$ (see Sect.~\ref{sec:proba}) and the corresponding mean first-passage times (MFPTs) $\tau_z$ (see Sect.~\ref{sec:mfpt}), both obtained with TAMS.
These quantities were estimated at $100$ intermediate levels $z$ of the degradation function, from $z=0.01$ to $z=1$.
Both quantities are complementary in describing the transition of the Amazon rainforest in regions 1 and 2 to a degraded state.
Figure~\ref{fig:proba_mfpt}a presents all values of $p_z$ for both studied regions, while Fig.~\ref{fig:proba_mfpt}b presents all MFPTs $\tau_z$ for the same regions.
Region $1$ is always presented in red and region $2$ in blue.
Note that both regions 1 and 2 have a nonzero MFPT across the level $z=0$ because the initial mean tree cover (from ERA5, shown in Fig.~\ref{fig:regions}) is larger than the equilibrium mean tree cover (taken as the baseline for the degradation function $\varphi$).
So all trajectories in both regions are initialized from a negative degradation function and need some time to reach the level $z=0$.

Let us first focus on region 2, where Fig.~\ref{fig:proba_mfpt}a shows that the transition to a degraded forest is inevitable: the probability of reaching any level $z$ of the degradation function within $t_{\max}$ is always one.
TAMS is initialized (see Sect.~\ref{sec:algo}) with a pool of $N$ trajectories, simulated until $t_{\max}$ or until they reach $z=1$.
When estimating a probability with TAMS, the only way to obtain a probability of one is to have all trajectories in this initial pool reach $z=1$ before $t_{\max}$.
The algorithm therefore stops before performing its first discarding/cloning step, so it exerts no selection pressure on the trajectories.
In particular, the algorithm cannot select trajectories with a weakened AMOC and a spontaneous AMOC is extrmely unlikely to happen to all trajectories in all TAMS runs.
Indeed, the hosing strength $\overline{E_A}$ tuned with CESM puts the AMOC box model into a regime where the AMOC collapse is rare: we showed in~\cite{Jacques-Dumas2024b} that for the parameters used here, the AMOC has a probability of about $10^{-4}$ to collapse within $100$ years.
Moreover, since the Amazon dynamics have no effect on the AMOC dynamics in this model, we can expect the AMOC to simply follow its natural variability around its on-state throughout the whole Amazon transition in region 2.
But, we know (from the potentials in Fig.~\ref{fig:potentials}) that, in the absence of noise and when forced by a constant AMOC in its on-state, the AMOC-Amazon model possesses a stable equilibrium at $73\%$ tree cover.
So, if the inevitable transition of the Amazon in region 2 cannot be explained by a change in the AMOC strength, it is due to the stochastic forcing by wildfires, which is not taken into account in the potentials of Fig.~\ref{fig:potentials}.
Indeed, the MAP in region $2$ is small enough even with a strong AMOC to allow for intense fires: see the fire intensity against MAP in Fig.~\ref{fig:fire_intensity} and the values of MAP in region 2 against the AMOC strength in Fig.~\ref{fig:map_mcwd}b.
In the tree cover model (Eq.~\ref{eq:amazon_model}), the diffusion term allows treeless regions to spread, and the MAP and MCWD terms are not large enough to regrow the tree cover.
As shown in Fig.~\ref{fig:regions}a, an AMOC collapse would increase the MAP in region $2$ and thus decrease the intensity of wildfires.
Figure~\ref{fig:proba_mfpt}b shows that the Amazon in region $2$ reaches its degraded state within $68$ years, so it makes an already unlikely AMOC collapse even more unlikely to occur before the rainforest finishes its transition.
However, tree cover may regrow, and the rainforest state might be restored if the simulation were run for longer and the AMOC eventually collapsed.

On the other hand, the transition of the rainforest to a degraded forest in region 1 is unlikely, with a transition probability of $5\times10^{-5}$, to be read for $z=1$ in Fig.~\ref{fig:proba_mfpt}a.
The transition occurs in four main stages.
First, from $z=0$ to $z=0.1$, the probability $p_z$ decreases from $1$ to $0.6$.
So, there is a $40\%$ chance that the Amazon cannot reach level $z=0.1$ of the degradation function before $t_{\max}$.
Moreover, Fig.~\ref{fig:proba_mfpt}b shows that the MFPTs $\tau_z$ increase very fast between $z=0$ and $z=0.1$: the system takes over $80$ years between the first crossing of the level $z=0$ and the first crossing of the level $z=0.1$.
Note that TAMS is constrained by the time horizon $t_{\max}$, so the algorithm ultimately selects the trajectories where the level $z$ of the degradation function is reached within $200$ years.
Yet, these selected trajectories still take $100$ years to make the first $10\%$ of the transition (i.e. reach level $z=0.1$ while the full transition corresponds to reaching $z=1$).
The drop in $p_z$ and the increase in $\tau_z$ mean that the events that initiate the degradation in region 1 take a long time to develop and are unlikely to occur within $t_{\max}=200$ years.
These results can be interpreted as the signature of large wildfires: in region $1$, MAP and MCWD are quite large in the AMOC on-state, so it is difficult for wildfires to develop.
They are unlikely to destroy a large portion of the forest and need time to perturb the forest out of its equilibrium.
It is therefore difficult for the tree cover to decrease and thus for the degradation function to increase.

Then, from $z=0.1$ to $z=0.5$ the probabilities $p_z$ decrease at a slower pace, from $0.6$ to $0.3$.
A slow change in $p_z$ means that the probability of reaching any level $z$ between $0.1$ and $0.5$ is comparable: once a level $z$ has been reached, it is not much more unlikely to reach another level $z'>z$.
So, after initial wildfires have developed, the system is easily destabilized because of the spread of the treeless areas.
Then, between $z=0.5$ and $z=0.9$, $p_z$ undergoes a much faster decrease as $z$ increases.
After reaching a degradation level of $0.5$, the steep decrease in $p_z$ indicates that reaching each level $z$ is increasingly more unlikely than reaching the former level.
So, the forest can no longer be easily destabilized by wildfires, and the transition is increasingly difficult.
In other words, at this stage, further tree cover degradation is only possible through very unlikely extreme events.

However, Fig.~\ref{fig:proba_mfpt}b, shows that the MFPT in region 1 increases slowly and steadily after the system has reached $z=0.1$, until reaching a degraded forest state.
Region 1 takes on average as much time to make the initial $10\%$ of the transition as to undergo the remaining $90\%$, although the second half of the transition is much more difficult than the first half ($p_z$ drops for $z>0.5$).
Therefore, tree cover loss greatly accelerates after initial wildfires have perturbed the forest out of equilibrium.
There is no inconsistency between the acceleration of tree cover loss and the fact that it is increasingly unlikely: extreme wildfires are very unlikely to occur, but they destroy tree cover fast.
So, on average, TAMS only selects trajectories where extreme wildfires occur, because it is the only way to make the forest transition within the time horizon $t_{\max}$.
Finally, the slope of $p_z$ between $z=0.9$ and $z=1$ is flat, although it is insignificant compared to the size of the $95\%$ confidence interval of $p_z$.

In summary, Fig.~\ref{fig:proba_mfpt} presents in each panel two complementary quantities to gain insight into the dynamics of the transition of the rainforest: the probability of reaching every level $z$ of the degradation function before $t_{\max}$ (panel a) and the mean time when these crossings first occur (panel b).
In region 1, the transition to a degraded forest occurs in several stages.
First, wildfires need a long time to develop because of the large initial values of MAP and MCWD.
It is relatively difficult to perturb the forest in region 1 out of equilibrium.
Past the initial perturbation, wildfires can spread more easily and further decrease tree cover.
However, the transition is still unlikely to occur within $200$ years because MAP and MCWD may remain too large to allow the rapid development of sufficiently extreme wildfires.
On the other hand, in region 2, the forest quickly transitions to its degraded state because of the presence of stochastic wildfires.
In the next section, we get additional insight into the dynamics of MAP and MCWD by analyzing the AMOC strength during the Amazon transition to a degraded forest.

\subsection{Tracked AMOC strength}
\label{sec:amoc_impact}

\begin{figure}[t]
	\centering
	\includegraphics[width=\textwidth]{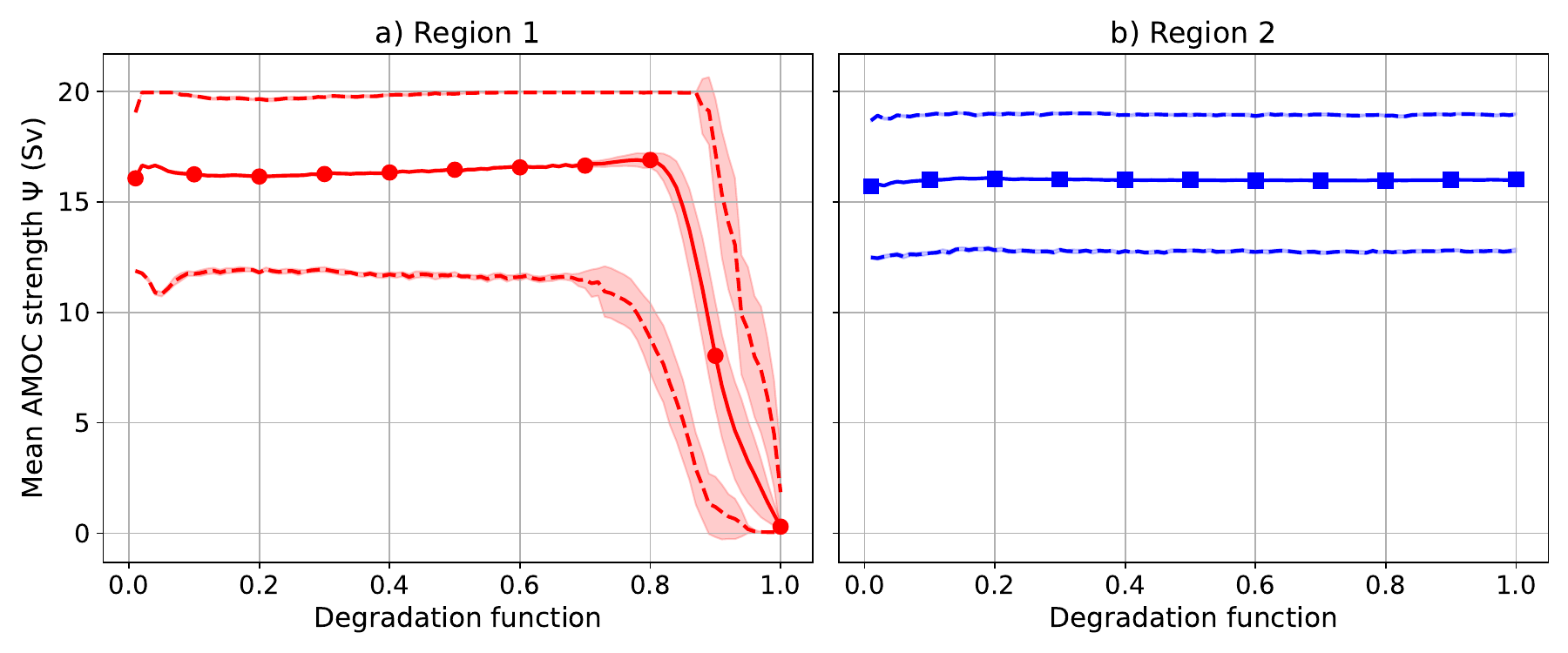}
	\caption{Mean AMOC strength $\Psi$ tracked across every level $z$ of the degradation function for the two regions defined in Fig.~\ref{fig:regions}. The solid line is the average of $\Psi$ over $20$ independent TAMS runs. Dashed lines show the $5^\mathrm{th}$ and $95^\mathrm{th}$ quantiles of the marginal distribution of all AMOC strengths at every level. The mean and the quantiles are all estimated at every run of TAMS, so the shaded areas represent their $95\%$ confidence intervals computed over the same runs. Markers indicate the tenths of the degradation function.}
	\label{fig:mean_amoc_strength}
\end{figure}

After describing the probability that the Amazon transitions in regions 1 and 2, we now turn to the dynamics of the AMOC.
As explained in Sect.~\ref{sec:distrib}, we can analyze how the AMOC behaves as the Amazon rainforest transitions to a degraded forest by estimating the distribution of AMOC strengths at every stage of the Amazon transition.
Let us stress that TAMS selects trajectories based on their mean tree cover only, to push them to a degraded state.
The algorithm is therefore totally blind to the AMOC component of these trajectories.
By analyzing the distribution of AMOC strength in the trajectories selected by TAMS, we can thus determine which values of AMOC strengths most favor the Amazon transition.
This information provides insight into the relation between both systems and into the impact of the AMOC on the Amazon rainforest in each selected region.
Figure~\ref{fig:mean_amoc_strength} presents the AMOC strength $\Psi$ tracked across the levels $z$ of the degradation function, conditioned on the fact that the Amazon rainforest collapsed within $200$ years.
Each run of TAMS estimates the whole distribution of $\Psi$ at every stage of the Amazon transition in regions 1 and 2.
In both panels of Fig.~\ref{fig:mean_amoc_strength}, the solid lines represent the mean of the distribution of $\Psi$, averaged over $20$ independent TAMS runs.
The dashed lines represent the $5^\mathrm{th}$ and $95^\mathrm{th}$ percentiles of $\Psi$, averaged over the same TAMS runs.

Figure~\ref{fig:mean_amoc_strength}a shows, at every stage of the Amazon transition, the distribution of the AMOC strength in region 1, where the AMOC systematically collapses ($\Psi\leq0$~Sv) as the forest reaches its degraded state.
Such systematic collapse suggests that the transition cannot occur without the drying, and increase in fire intensity, following an AMOC collapse.
So, the AMOC must collapse for the Amazon rainforest in region 1 to collapse.
But the AMOC starts weakening only after the degradation function has reached $z=0.8$.
Before that level, i.e. during $80\%$ of the Amazon transition in region 1, the AMOC strength and its $5^\mathrm{th}$ and $95^\mathrm{th}$ remain constant on average.
In other words, TAMS does not particularly select the trajectories exhibiting an AMOC weakening to maximize the degradation in region 1.
Before the degradation function reaches $z=0.8$, wildfires alone are sufficient to destroy the forest, and do not require region 1 to dry because of an AMOC collapse (see Fig.~\ref{fig:regions}a).
Every iteration of TAMS discards the trajectories having the smallest degradation.
Up to $z=0.8$, trajectories with a wide range of AMOC strengths can exhibit an increase in degradation, so the selection pressure exerted by TAMS does not introduce any statistical bias in the distribution of AMOC strength.
However, when the degradation function reaches the level $z=0.8$, the AMOC strength suddenly drops.
In other words, once the system reaches $z=0.8$, the degradation function can only increase in trajectories where the AMOC collapses and triggers a drying of region 1, thus more intense fires.
Therefore, iteration after iteration, TAMS discards all trajectories where the AMOC strength remains constant, or does not decrease fast enough.
We can read in Fig.~\ref{fig:proba_mfpt}b that only a couple of years separate the MFPT across level $z=0.8$ and the MFPT across level $z=1$.
This very short time scale should make such sudden AMOC collapse extremely unlikely: as shown in~\cite{Jacques-Dumas2024b}, for the model parameters taken here, the AMOC has a probability of about $10^{-4}$ to collapse within $100$ years, and this event takes about $60$ years.
Such a fast AMOC strength decrease within a few years is therefore only due to a strong selection by TAMS of the trajectories having the largest degradation, which correlates with a low AMOC strength.

Now, we focus on the panel b of Fig.~\ref{fig:mean_amoc_strength}, which presents the distribution of the AMOC strength in region 2 during the Amazon transition.
The average AMOC strength and its $5^\mathrm{th}$ and $95^\mathrm{th}$ percentiles all remain approximately constant during the whole transition of region 2 to a degraded forest.
This is consistent with the fact that intense wildfires trigger a very fast transition to a degraded forest, regardless of the AMOC strength.

We can conclude from Fig.~\ref{fig:mean_amoc_strength}a that TAMS only selects trajectories exhibiting an AMOC collapse as the forest in region 1 approaches its degraded state.
It suggests that the Amazon rainforest cannot transition if the AMOC does not collapse.
In region 2, wildfires cause the Amazon to transition fast, so no weakening of the AMOC is needed to favor the transition.
As a result, the average AMOC strength remains constant during the transition, as shown in Fig.~\ref{fig:mean_amoc_strength}b.

In Sect.~\ref{sec:cesm}, we have derived, for regions 1 and 2, a simple expression of MAP and MCWD as a function of latitude and AMOC strength.
Figure~\ref{fig:mean_amoc_strength} presents the estimated mean AMOC strength at every stage of the transition of the forest in these regions to a degraded forest state within $200$ years.
The corresponding mean MAP and MCWD in both regions can then be reconstructed as a function of latitude and degradation level.
In Appendix~\ref{app:reconstructed}, Fig.~\ref{fig:rebuilt_map} and Fig.~\ref{fig:rebuilt_mcwd} respectively show the reconstructed mean MAP and MCWD, based on the estimated mean AMOC strength from Fig.~\ref{fig:mean_amoc_strength}.
For region 1, where the AMOC suddenly collapses after the degradation function reaches $z=0.8$, these reconstructions reflect the relationship between MAP, MCWD and the AMOC strength shown in Fig.~\ref{fig:map_mcwd_states} (also in Appendix~\ref{app:reconstructed}) for the AMOC on- and off-states.
In region 2, the mean MAP and MCWD both remain constant during the transition to a degraded forest, consistent with the fact that the mean AMOC strength also remains constant.
Moreover, we also show in the Appendix~\ref{app:reconstructed} (Fig.~\ref{fig:rebuilt_fire_intensity}) the evolution of the mean fire intensity $f(T,P)$ in regions 1 and 2 as a function of latitude and degradation.
This reconstruction is based on the values of MAP presented in Fig.~\ref{fig:rebuilt_map} and on the mean tree cover corresponding to every level of the degradation function.

Fire intensity increases in region 1, and in particular in the north, where it was initially very small due to a large MAP.
Furthermore, the fire intensity increases the most rapidly when the mean tree cover approaches $0.57$, which is close to the degraded forest state, which the system reaches when the degradation function equals one.
So, as the forest in region 1 reaches the threshold to a degraded state, the strong increase in fire intensity is explained by the combination of two effects.
First, the strong AMOC weakening triggers an decrease in MAP, which increases the fire intensity.
Second, the fire intensity is also increased by the decrease in tree cover, which allows a better propagation of the flames.
In region 2, the fire intensity also increases because the mean tree cover decreases, although MAP remains constant.

\subsection{Cascading probability}

\begin{figure}[t]
	\centering
	\includegraphics[width=\textwidth]{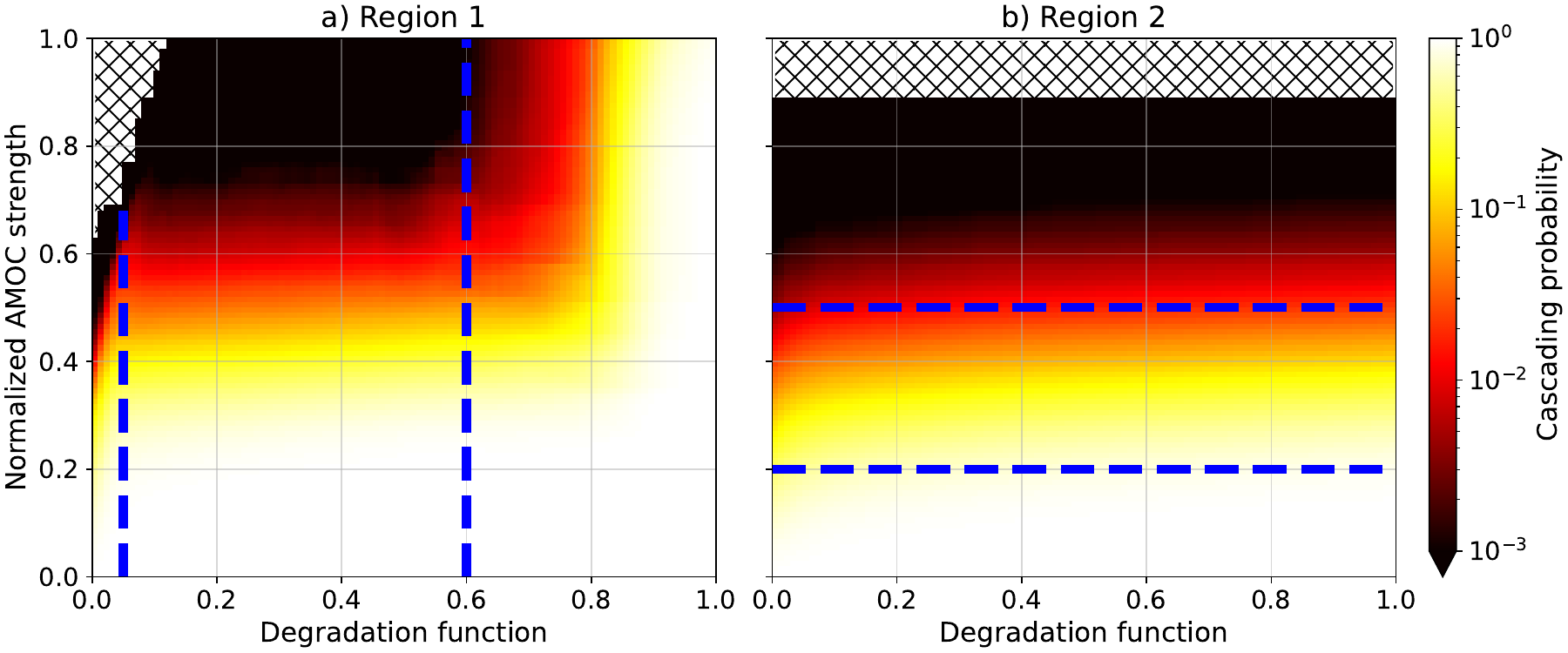}
	\caption{Probability that any decrease in the AMOC strength occurs before any increase in the degradation of the Amazon rainforest, given that the latter occurs within $200$ years, for both regions shown in Fig.~\ref{fig:regions}. The x-axis represents the degradation function, while the y-axis shows the normalized AMOC strength. Probabilities are clipped below $10^{-3}$ and the hatched areas correspond to zero values. Blue dashed lines correspond to values of the degradation function (in panel a) or the normalized AMOC strength (in panel b) discussed in the text.}
	\label{fig:cascading_probability}
\end{figure}

After separately analyzing the dynamics of the transition of the Amazon rainforest and of the AMOC collapse in regions 1 and 2, we now focus on the interaction between both systems.
Here, we try to determine the likelihood of a transition of the Amazon rainforest within $t_{\max}$ as a consequence of an AMOC weakening.
To that end, we compute the probability that any decrease in AMOC strength occurs before any tree cover loss in the Amazon rainforest, given that this tree cover loss occurs within $t_{\max}=200$ years.
For readability, here, we do not use the AMOC strength as is.
Instead, we normalize it so that it is equal to zero when the AMOC is in its on state ($\Psi=\Psi_{\mathrm{ON}}$) and to one when it has collapsed ($\Psi=0$~Sv).
More precisely, the normalized AMOC strength along trajectory $\mathbf{X}$ reads: $1-\frac{\Psi(\mathbf{X})}{\Psi_\mathrm{ON}}$.
The conditional probability $\mathbb{P}(\tau_\psi<\tau_z|\tau_z<t_{\max})$ (see Eq.~\ref{eq:full_cond}) is presented in Fig.~\ref{fig:cascading_probability} for all levels $z$ of the degradation function (on the x-axis) and for all levels $\psi$ of the normalized AMOC strength (on the y-axis).

Each grid cell $(z,\psi)$ in both panels of Fig.~\ref{fig:cascading_probability} gives the probability that a decrease in the AMOC strength by $100\times\psi\%$ occurs before an increase in the Amazon degradation by $100\times z\%$, given that the latter occurs within $200$ years.
Hatched cells correspond to a probability of zero: the corresponding event is so unlikely that it was never sampled in the trajectories generated through TAMS.
The upper right-hand corner of both panels gives the conditional probability that the AMOC reaches a normalized strength of one (i.e. reaches its collapsed state) before the Amazon rainforest in region 1 or 2 reaches a degradation of one (i.e. reaches a degraded state).
Therefore, the upper right-hand corner of each panel gives the probability that the AMOC collapses before the Amazon rainforest transitions (in the corresponding region), given that the forest transitions within $200$ years.
Similarly, the upper left-hand corner of every panel gives the probability that the AMOC collapses before the Amazon has suffered any degradation.
This event is so unlikely that it never occurred in all trajectories simulated throughout the $20$ independent runs of TAMS.

In region 2 (Fig.~\ref{fig:cascading_probability}b), the probability that the AMOC weakens before the degradation increases is approximately independent of the degradation function: it does not evolve during the transition of the Amazon rainforest.
This is consistent with Fig.~\ref{fig:mean_amoc_strength}b, where we found that the mean AMOC strength remains constant throughout the Amazon transition, which occurs on a smaller time scale than that of the AMOC collapse.
But the AMOC is still subject to its natural variability, so it may still weaken during the Amazon transition.
For instance, there is a probability of about $80\%$ that the normalized AMOC strength reaches $0.2$ (lower blue line in Fig.~\ref{fig:cascading_probability}b) before the Amazon reaches any level of the degradation function.
So, the AMOC is ensured to weaken by $20\%$ at some stage during the Amazon transition.
However, it is very unlikely that the AMOC weakens significantly during the Amazon transition: the AMOC strength has a probability of only $0.02$ to decrease by $50\%$ (upper blue line in Fig.~\ref{fig:cascading_probability}b).

In region 1 (Fig.~\ref{fig:cascading_probability}a), the probability that the AMOC collapses before the Amazon reaches its degraded state is also consistent with the results in Fig.~\ref{fig:mean_amoc_strength}.
Indeed, for a degradation function $z>0.8$, the cascading probability approaches one for a normalized AMOC strength of one.
So, after the degradation function has crossed $z=0.8$, there is a probability one that the AMOC reaches its collapsed state before the degradation function increases further.
But the likelihood of an AMOC weakening already increases earlier in the Amazon transition.
When the degradation function reaches $z=0.6$ (rightmost blue line in Fig.~\ref{fig:cascading_probability}a), $\mathbb{P}(\tau_\psi<\tau_z|\tau_z<t_{\max})$ starts increasing for all normalized AMOC strengths $\psi\geq0.7$.
If a trajectory reaches a degradation level of $0.6$, there is only a $10^{-3}$ probability that the AMOC has already weakened by $80\%$ to $100\%$ before $z=0.6$ is reached.
But the probability to observe such AMOC weakening before reaching a degradation level of $0.7$ is multiplied by $10$.
Therefore, the likelihood of observing an AMOC collapse rapidly increases when the Amazon rainforest has been degraded by $60\%$, whereas the probability of any AMOC weakening had remained constant for degradation levels between $0.05$ and $0.6$.

We stress that we have not forced the AMOC towards a collapse: TAMS selects trajectories to maximize the degradation function of the ensemble.
From the moment when the degradation function reaches $0.6$ to $0.7$, all selected trajectories exhibit a strong AMOC weakening.
The conditional probabilities shown in Fig.~\ref{fig:cascading_probability} are estimated from the statistics measured in these selected trajectories, so the selection performed by TAMS results in an increase in the probability that an AMOC collapse occurs before crossing larger degradation levels.
That same selection phenomenon can be seen from another viewpoint: trajectories that do not exhibit an AMOC collapse, or a strong weakening, are not selected.
So, these discarded trajectories are not able to maximize their degradation as efficiently as the trajectories exhibiting an AMOC collapse.
The selection performed by TAMS, and the increase in the conditional probability, suggest that an AMOC collapse is a necessary condition for the degradation function to reach levels larger than $z=0.7$.

Finally, Fig.~\ref{fig:cascading_probability}a also shows that, very early in the Amazon transition in region 1, a weaker AMOC also favors degradation.
The leftmost blue line in Fig.~\ref{fig:cascading_probability}a indicates an increase by one order of magnitude of the probability that the AMOC weakens by $20\%$ to $60\%$ before the degradation function reaches $z=0.05$, compared to that same AMOC weakening occurring before any degradation of the rainforest ($z=0$).
Such increase in the probability of an AMOC weakening between $z=0$ and $z=0.05$ shows that, already early in the Amazon transition, TAMS selects trajectories where the AMOC strength decreases to help wildfires perturb the forest out of equilibrium through a decrease in MAP.
This selection by TAMS of trajectories exhibiting such AMOC weakening, which is an unlikely process that may take time to develop is consistent with our analysis of Fig.~\ref{fig:proba_mfpt}.
There, we found indeed that the probability that the degradation function reaches $z=0.05$ before $t_{\max}$ is only $60\%$ and region 1 takes $100$ years on average to reach that level.
So, the wildfires that degrade the rainforest can only be made more likely by the decrease in MAP triggered by an AMOC weakening, which is therefore selected by TAMS.

In summary, Fig.~\ref{fig:cascading_probability} shows that, at several stages of the transition of the Amazon rainforest to a degraded forest state in region 1, TAMS systematically selects trajectories with a weaker AMOC, in particular towards the end of the Amazon transition.
Here, the AMOC weakens significantly whenever there is a drop in the probability of reaching larger levels of the degradation function: at the beginning of the Amazon transition, and after reaching $z=0.6$.
Moreover, there is a probability of one that the AMOC strongly weakens and collapses before the forest in region 1 reaches its degraded state.
These elements together suggest that the AMOC collapse is not merely correlated to the Amazon transition, but is a necessary condition for that transition to take place.

\section{Discussion and outlook}
\label{sec:discussion}

We have applied the rare-event algorithm TAMS to a coupled conceptual model of the AMOC and the Amazon rainforest.
It allowed us to estimate the probability that the Amazon rainforest transitions to a degraded forest state within $200$ years in two different regions.
We then obtained at every stage of tree cover loss: the mean first-passage time, the mean AMOC strength, and the reconstructed MAP, MCWD and mean fire intensity.
Finally, we estimated the conditional probability that the AMOC collapses before the rainforest does, given that the latter collapses within $200$ years.
We were able to derive a range of such cascading factors, quantifying the influence of any decrease in the AMOC strength on any level of degradation of the Amazon rainforest.

For two coupled systems $A$ and $B$, the cascading probability $\mathbb{P}(\tau_A<\tau_B\ |\ \tau_B<t_{\max})$ is easy to interpret.
A value of zero means that the collapse of $A$ never occurs before the collapse of $B$, given that the latter occurs before $t_{\max}$: either $A$ cannot collapse faster than $B$ or earlier than $t_{\max}$, or the collapse of $A$ hinders that of $B$.
In region 2, the cascading probability has a value of zero because the transition of the Amazon rainforest occurs too fast, without requiring an AMOC weakening, although Fig.~\ref{fig:regions}b shows that an AMOC collapse would make region 2 wetter.
A value of $0.5$ means that a collapse of $A$ may or may not occur before that of $B$ and before $t_{\max}$, indicating the absence of a causal link between the transition of both systems.
Finally, a value of one of the cascading probability means that the collapse of $A$ always occurs before that of $B$.
In the conceptual AMOC model, \cite{Jacques-Dumas2024b} showed that an AMOC collapse is unlikely in the setup chosen here, so a probability of one suggests that the tipping of $A$ is a necessary condition for the tipping of $B$.
It is the case in region 1, where the drying due to an AMOC collapse is necessary to trigger intense enough forest fires and bring the Amazon rainforest to a degraded forest.
The presence of a causal link in this case is suggested by the fact that an AMOC weakening is \textit{systematically} selected by TAMS to degrade the mean tree cover in region 1.
The causal link hypothesis is reinforced by the careful analysis of both systems through other quantities estimated by TAMS.
In particular, the probability that any weakening of the AMOC strength occurs before any degradation shows that the rainforest in region 1 cannot transition to a degraded state without being preceded by an AMOC collapse.

Although we use here a conceptual model of the AMOC and the Amazon rainforest, it is tuned to the output of CESM, a General Circulation Model (GCM).
We show that, based on CESM data, an AMOC collapse may endanger the rainforest by causing the northwest of Brazil (region 1) to dry.
It is therefore important to consider an AMOC collapse as a source of instability for the rainforest, in connection with global warming and deforestation.
The drying of the northwest of Brazil in the case of an AMOC collapse is consistent with the findings of~\cite{Parsons2014}, \cite{Nian2023} and \cite{benYami2024}.
However, \cite{Nian2023} also considered the cooling effect of an AMOC collapse and found that it may stabilize the rainforest by opposing the effect of global warming.
The cascading tipping between the AMOC and the Amazon rainforest has also been studied in non-process-based models, where both systems are modeled as a double-well potential system which are linearly coupled~\cite{Ciemer2021,Wunderling2021,Wunderling2023}.
Consistent with~\cite{Nian2023}, \cite{Ciemer2021} found that an AMOC collapse may stabilize the rainforest and prevent its collapse whatever the increase in global mean temperature.
The findings of these studies are not directly comparable to our findings, since we did not consider global warming, but only the drying effect caused by an AMOC collapse.
However, all tipping cascade studies rely on a Monte-Carlo sampling of the collapse of the AMOC and the Amazon rainforest to determine their probabilities.
Here, we estimate these probabilities using TAMS, which is much more efficient than Monte-Carlo when probabilities are small~\cite{Brehier2015b}.
But, more importantly, TAMS can sample a wide variety of functions of the system trajectories, with better mathematical properties than Monte-Carlo sampling.
Beyond transition probabilities, TAMS thus allows a detailed analysis of any property of the studied system.
Like previous studies, we studied here a coupled conceptual model of the AMOC-Amazon system, but the originality of the present work lies in the method used to extract information from such a model.
We designed a process-based model from CESM data to show that TAMS allows extracting physical insight from the sampled transitions.

Our study should of course be seen as a proof-of-concept rather than a quantitative demonstration of the effect of an AMOC weakening on the Amazon rainforest.
First, the precipitation biases in CESM over the Amazon~\cite{Sakaguchi2018} affect both MAP and MCWD.
Moreover, the modeled AMOC dynamics in CESM are biased as well when compared to reanalysis~\cite{vanWesten2024c}, which influences the tuning of the conceptual AMOC model as well as the precipitation patterns over the Amazon.
Our conceptual Amazon model is also very simple; reducing the Amazon rainforest to a single 1D tree cover is an obvious oversimplification.
One of the main effects of an AMOC collapse on the Amazon would be the shift in the seasonal cycle and the monsoon~\cite{benYami2024}, which is not accounted for here.
\cite{Wunderling2025} have recently shown that large-scale teleconnections between Amazon regions through evapotranspiration may play a much larger role than MAP and MCWD in rainforest tipping.
Moreover, \cite{Flores2024} have pinpointed global warming and deforestation as two of the main drivers of stability loss for the Amazon rainforest, and an extended conceptual model should also include these forcings.
We attempted here to keep the Amazon model as simple as possible while connecting it to AMOC dynamics to demonstrate the power of our quantitative approach regarding tipping cascades.
It would be very interesting as a next step to couple more precise Amazon models (e.g.~\cite{Wuyts2017}) to an AMOC model to precisely quantify the dependence of the Amazon on the AMOC.

However, all these improved models remain conceptual, compared to directly studying General Circulation Models (GCMs), where all climate subsystems are already coupled.
Moreover, such a study would no longer be limited to the mere one-sided influence of the AMOC on the Amazon rainforest: we would be able to quantify the impact of any climate subsystem on the Amazon.
Indeed, in a GCM or even in an Earth Model of Intermediate Complexity (EMIC), the cost of simulating trajectories would always remain orders of magnitude larger than that of estimating any number of functions of the system trajectories (all the more so that the estimation step of TAMS can be run in parallel during the simulation of trajectories for the next iteration).
It is currently not possible to apply TAMS to such large and complex systems, due to the prohibitive computational cost of simulating hundreds of trajectories, with many restarts.
Moreover, TAMS would have to be run several times to obtain statistics on the estimated functions.
But TAMS is not limited to conceptual models and could currently be applied to systems of much larger dimension, with up to $10^5$ variables.
To obtain quantitative estimates of the probabilities of tipping cascades, directly relatable to the Earth system, a possibility would be to apply TAMS to a large emulator of several tipping elements, tuned to a state-of-the-art climate model.

Another possibility would be to use another rare-event algorithm, such as Giardina-Kuchan-Tailleur-Lecomte (GKTL)~\cite{Lestang2018,Ragone2018}.
It has already been applied to PlaSim~\cite{Ragone2018,Cini2024}, which lies in the category of EMICs, and even to CESM~\cite{Ragone2021}.
The main advantage of GKTL over TAMS is that the ensemble of trajectories is only run once, so the cost of GKTL is exactly that of initializing TAMS.
However, TAMS is more flexible in the range of possible experiments and seems to have a lower variance~\cite{Lestang2018}.
Moreover, GKTL is heavily dependent on a few parameters that have to be empirically tuned~\cite{Ragone2018} and may greatly affect its efficiency.
Although GKTL could, in principle, be used to quantify cascading tipping probabilities as done here, it has yet never been applied to such a complex case.
Because of its much cheaper computational cost, it would be very interesting to adapt GKTL to our framework and apply it to tipping cascades in an EMIC or a GCM.

Another step beyond the present work would be, instead of conditioning all results to a collapse of the Amazon within $200$ years, to compute the probability $\mathbb{P}(\tau_\mathrm{AMOC}<\tau_\mathrm{Amazon})$.
This is not feasible with direct numerical simulation because it would be prohibitively expensive to wait for the occurrence of two rare events in the right order.
Moreover, TAMS does not solve this problem either because trajectories would only be stopped after either the AMOC or the Amazon rainforest has collapsed, which might still be prohibitively expensive, even in a simple model, depending on the model parameters.
However, this problem could be overcome by taking inspiration from ancestor algorithms of TAMS, such as Multilevel Splitting~\cite{Glasserman1998,Glasserman1999,Garvels2002}.
The distance to the collapse of both systems could be measured using two score functions, one for each system.
The algorithm would then stop trajectories as soon as they reach the next gap of either score function.
In this particular case, older algorithms might be more efficient than TAMS and GKTL, but they have never been applied to such complex setups.

We have shown in our study that the drying of the Amazon basin caused by an AMOC collapse may endanger the rainforest by increasing the intensity of wildfires.
More importantly, we showed that TAMS is a valuable tool to analyze in-depth the relation between tipping elements by efficiently extracting information from trajectories simulated in a coupled model.

\clearpage

\section*{Acknowledgments}
The authors wish to thank Arie Staal for his helpful guidance in the design of the conceptual Amazon model and for providing the data used to make Fig.~\ref{fig:potentials}.
They also thank René van Westen for answering any AMOC-related questions, providing the CESM data and accompanying codes to compute the MAP and MCWD and relate it to AMOC strength.
This project has received funding from the European Union's Horizon~2020 research and innovation program under the Marie Sklodowska-Curie grant agreement no.~956170.
H.~A.~Dijkstra also received funding from the European Research Council through the ERC-AdG project TAOC (project~101055096, PI: Dijkstra).

\section*{Conflict of Interest}
The authors have no conflicts to disclose.

\section*{Competing interests}
The authors have no competing interest.

\section*{Author Contributions}
\textbf{Valerian Jacques-Dumas}: conceptualization (lead), software (lead), formal analysis (lead), writing - original draft (lead), writing - review and editing (equal). \textbf{Henk A. Dijkstra}: conceptualization (supporting), formal analysis (supporting), writing - original draft (supporting), writing - review and editing (equal).

\section*{Data Availability Statement}
The Python implementation of both the Amazon and the AMOC model, their coupling, the TAMS framework, the code producing figures and the output files can be found at the following address:
https://doi.org/10.5281/zenodo.15776754 \cite{code}.

\newpage

\bibliographystyle{plain}
\bibliography{biblio.bib}

\newpage

\appendix

\section{Tree cover potentials for different values of $s$}
\label{app:potentials}

\begin{figure}[h]
	\centering
	\includegraphics[width=\textwidth]{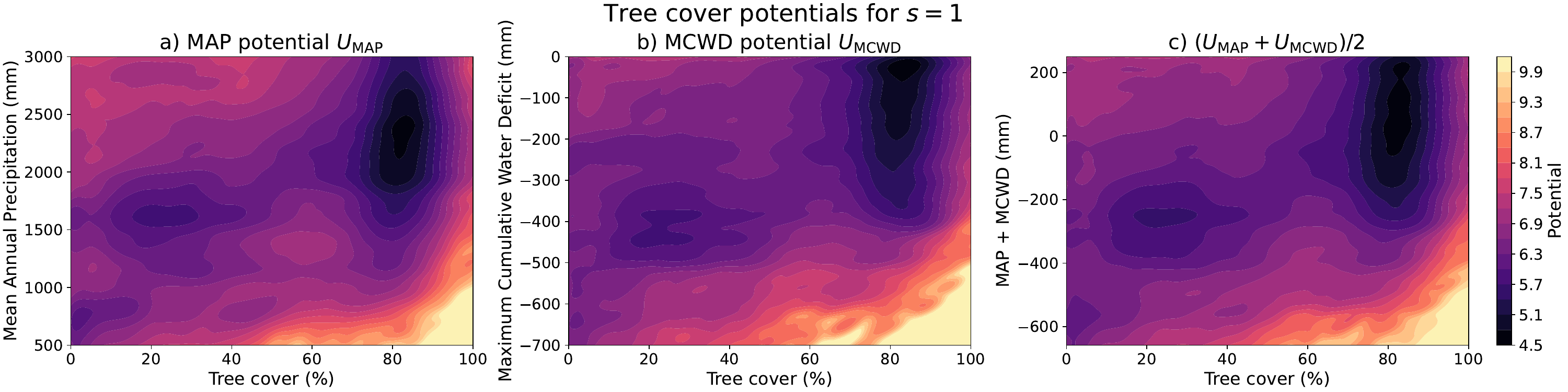}
	\caption{Potentials obtained from the procedure detailed in Sec.~\ref{sec:potentials} but with a bandwidth parameter $s=1$. In this case, the too narrow bandwidth creates spurious minima of the potential. The left panel shows the Mean Annual Precipitation (MAP) potential, the middle panel shows the Maximum Cumulative Water Deficit (MCWD) potential and the right panel shows their sum.}
	\label{fig:pot_1}
\end{figure}

\begin{figure}[h]
	\centering
	\includegraphics[width=\textwidth]{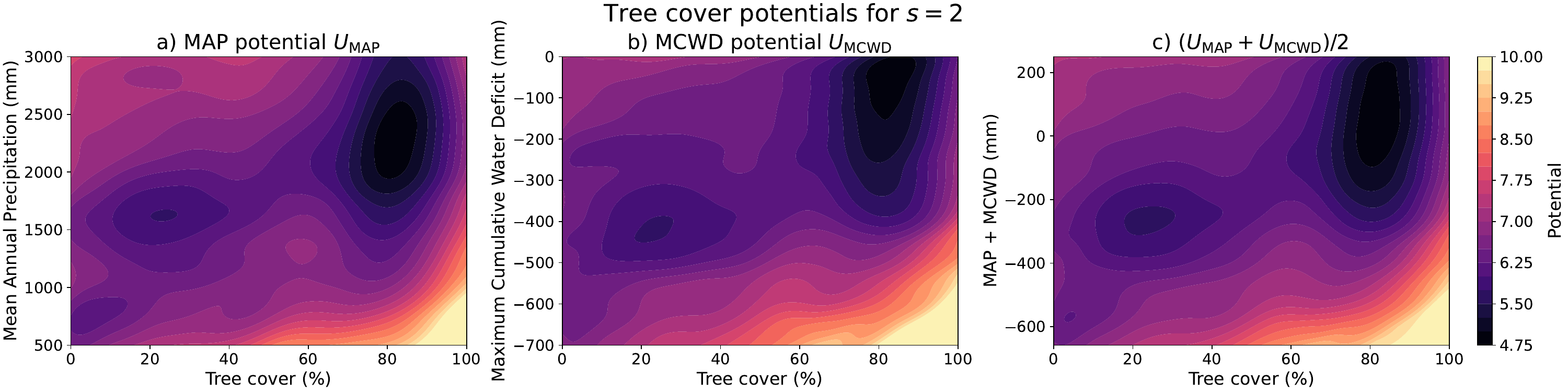}
	\caption{Potentials obtained from the procedure detailed in Sec.~\ref{sec:potentials} but with a bandwidth parameter $s=2$. In this case, the too narrow bandwidth creates spurious minima of the potential. The left panel shows the Mean Annual Precipitation (MAP) potential, the middle panel shows the Maximum Cumulative Water Deficit (MCWD) potential and the right panel shows their sum.}
	\label{fig:pot_2}
\end{figure}

\begin{figure}
	\centering
	\includegraphics[width=\textwidth]{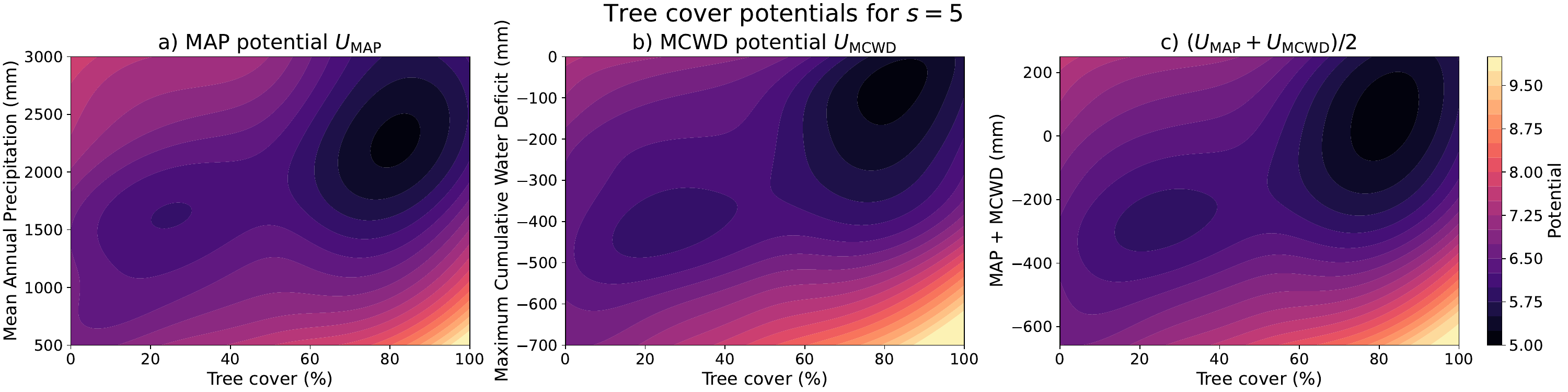}
	\caption{Potentials obtained from the procedure detailed in Sec.~\ref{sec:potentials} but with a bandwidth parameter $s=5$. In this case, the bandwidth is too large to allow for a clear rainforest / degraded forest bistability. The left panel shows the Mean Annual Precipitation (MAP) potential, the middle panel shows the Maximum Cumulative Water Deficit (MCWD) potential and the right panel shows their sum.}
	\label{fig:pot_5}
\end{figure}

\begin{figure}[h]
	\centering
	\includegraphics[width=\textwidth]{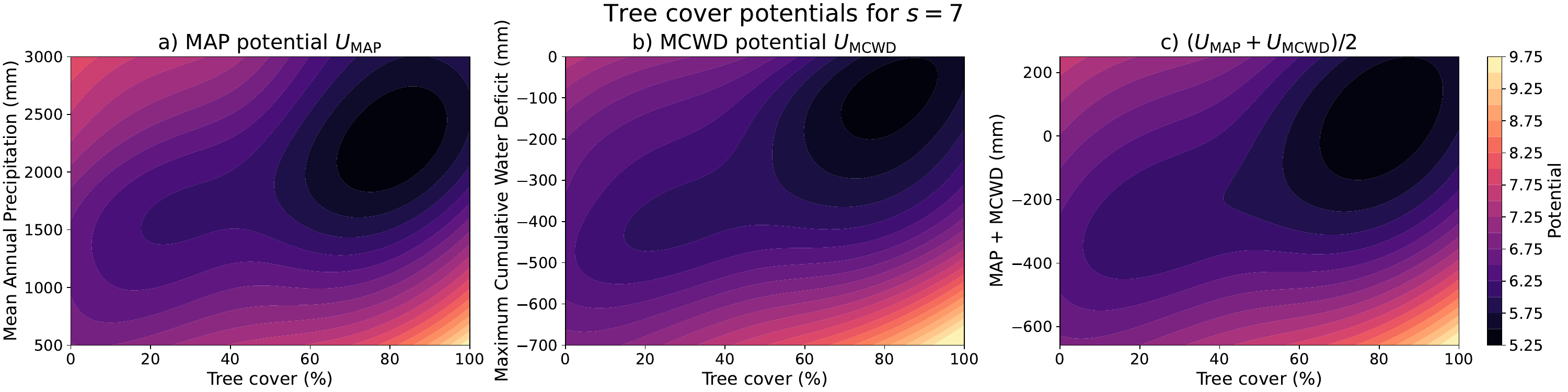}
	\caption{Potentials obtained from the procedure detailed in Sec.~\ref{sec:potentials} but with a bandwidth parameter $s=7$. In this case, the bandwidth is too large to allow for a clear rainforest / degraded forest bistability. The left panel shows the Mean Annual Precipitation (MAP) potential, the middle panel shows the Maximum Cumulative Water Deficit (MCWD) potential and the right panel shows their sum.}
	\label{fig:pot_7}
\end{figure}

\clearpage

\section{TAMS algorithm}
\label{app:tams}

TAMS drives an ensemble of $N$ trajectories from an initial set $A$ to a target set $B$ before a time horizon $t_{\max}$.
Here, $A$ is for each region a singleton (AMOC on state, Amazon rainforest initial state).
The set $B$ corresponds to the degraded  state of the Amazon rainforest, defined as a mean tree cover below $60\%$.
The main idea of the algorithm is to split the distance between $A$ and $B$ into a family of levels of the score function $\varphi$.
These levels are automatically determined at every iteration of the algorithm by the least successful trajectories, i.e. those having the lowest maximum value of the score function $\varphi$.
At every iteration, these least successful trajectories are discarded and replaced by branching other trajectories such that the ensemble is guaranteed to get closer to $B$.

To ensure the unbiasedness of the algorithm, the maximum level $z_{\max}$ of $\varphi$ beyond which the algorithm stops has to be defined such that $B\subset\{x\in\mathbb{R}^d\ |\ \varphi(x)>z_{\max}\}$.

The following description is heavily inspired by~\cite{Brehier2016}.

\paragraph{Notations}
\begin{itemize}
	\item At iteration $i$, the ensemble of $N$ members is denoted as $(\mathbf{X}^{(n,i)})_{1\leq n\leq N}$, with their corresponding weights $(W^{(n,i)})_{1\leq n\leq N}$.
	\item We denote as $L^{i}$ the set of labels of all trajectories computed until iteration $i$. Among this set, $L_\mathrm{on}^{(i+1)}$ refers to the trajectories retained for the next iteration and $L_\mathrm{off}^{(i+1)}$ to all those that have been discarded up to iteration $i$.
\end{itemize}

\paragraph{Initialization step}
\begin{itemize}
	\item Simulate $N$ trajectories $(\mathbf{X}^{(n,0)})_{1\leq n\leq N}$ starting from $A$ until they reach either $A$ (at time $\tau_A$) or time $t_{\max}$.
	\item Initialize the labels $L^{(0)}=\{1,\dots,N\}=L_\mathrm{on}^{(0)}$.
	\item Initialize the weight of each trajectory: $\forall n\in[1,N],\ W^{(n,0)} = 1/N$.
	\item Compute the score function $\varphi$ along each trajectory $\mathbf{X}^{(n,0)}$ and call $\Theta^{(n,0)}$ its corresponding maximum value.
	\item Sort $(\Theta^{(n,0)})_{1\leq n\leq N}$ in ascending order.
	\item Call $Z^{(0)}$ the $k$-th unique value in the sorted values of $\Theta^{(n,0)}$. If all values of $\Theta$ are inferior or equal to $Z^{(0)}$, set $Z^{(0)}=+\infty$. This case is called extinction.
	\item Set the number of iterations $i=0$.
\end{itemize}

\paragraph{Stopping criterion}
Stop the algorithm as soon as $Z^{(i)}>z_{\max}$. If this is the case, set the final number of iterations $I=i$. Otherwise, perform the next iteration.

\paragraph{Main loop}
\subparagraph{Splitting step}
\begin{itemize}
	\item The set $L_\mathrm{on}^{(i)}$ can be partitioned as:
$$L_\mathrm{on}^{(i)}=L_{\mathrm{on},\leq Z^{(i)}}^{(i)}\cup L_{\mathrm{on},>Z^{(i)}}^{(i)},$$
where the former contains all trajectories retained until now, but which value of $\Theta$ is inferior or equal to $Z^{(i)}$; the latter is defined similarly, but the values of $\Theta$ are strictly larger than $Z^{(i)}$.
	\item There are $K^{(i+1)}$ trajectories such that $\Theta^{(n,i)}\leq Z^{(i)}$. The branched trajectories to compute will be labeled with a new set of labels $L_\mathrm{new}^{(i+1)}=\{\mathrm{card}\ L^{(i)}+1,\dots,\mathrm{card}\ L^{(i)}+K^{(i+1)}\}$.
	\item The parents of the branched trajectories to compute are selected randomly within $L_{\mathrm{on},>Z^{(i)}}^{(i)}$.
	\item  Update the set of labels as follows:
$$L_\mathrm{on}^{(i+1)}=L_{\mathrm{on},>Z^{(i)}}^{(i)}\cup L_\mathrm{new}^{(i+1)},\
L_\mathrm{off}^{(i+1)}=L_\mathrm{off}^{(i)}\cup L_{\mathrm{on},\leq Z^{(i)}}^{(i)},\
L^{(i+1)}=L_{\mathrm{on}}^{(i+1)}\cup L_{\mathrm{off}}^{(i+1)}$$
	\item Update the weights as follows:
\begin{equation}
\begin{cases}
	W^{(n,i+1)}=W^{(n,i)}\ \forall n\in L_\mathrm{off}^{(i+1)}\\
	W^{(n,i+1)}=\frac{N-K^{(i+1)}}{N}W^{(n,i)}\ \forall n\in L_{\mathrm{on},>Z^{(i)}}^{(i)}\\
	W^{(n,i+1)}=\mathrm{weight\ of\ their\ parent\ replica\ after\ update}\ \forall n\in L_\mathrm{new}^{(i+1)}
\end{cases}
\end{equation}
\end{itemize}

\subparagraph{Resampling step}
\begin{itemize}
	\item For all values of $n$ belonging to $L^{(i)}$, trajectory $\mathbf{X}^{(n,i+1)}$ is the clone of trajectory $\mathbf{X}^{(n,i)}$.
	\item The trajectories $\mathbf{X}^{(n,i+1)}$ such that $n$ belongs to $L_\mathrm{new}^{(i+1)}$ are obtained by branching their parent at the time $\tau_{Z^{(i)}}=\min_{0\leq t\leq\min(\tau_A,t_{\max})}\{\varphi(\mathbf{X}(t))>Z^{(i)}\}$ and then independently simulated until time $\min(\tau_A,t_{\max})$.
\end{itemize}

\subparagraph{Level computation step}
\begin{itemize}
	\item Sort $(\Theta^{(n,i+1)})_{n\in L_\mathrm{on}^{(i+1)}}$ in ascending order.
	\item Set $Z^{(i+1)}$ the $k$-th unique value among the sorted $\Theta^{(n,i+1)}_{n\in L_\mathrm{on}^{(i+1)}}$. If all values of $\Theta^{(n,i+1)}_{n\in L_\mathrm{on}^{(i+1)}}$ are inferior or equal to $Z^{(i+1)}$, set $Z^{(i+1)}=+\infty$.
	\item Set the number of iterations $i=i+1$.
	\item Check the stopping criterion.
\end{itemize}

\newpage

\section{Reconstructed hydrological variables}
\label{app:reconstructed}

\begin{figure}[h]
	\centering
	\includegraphics[width=\textwidth]{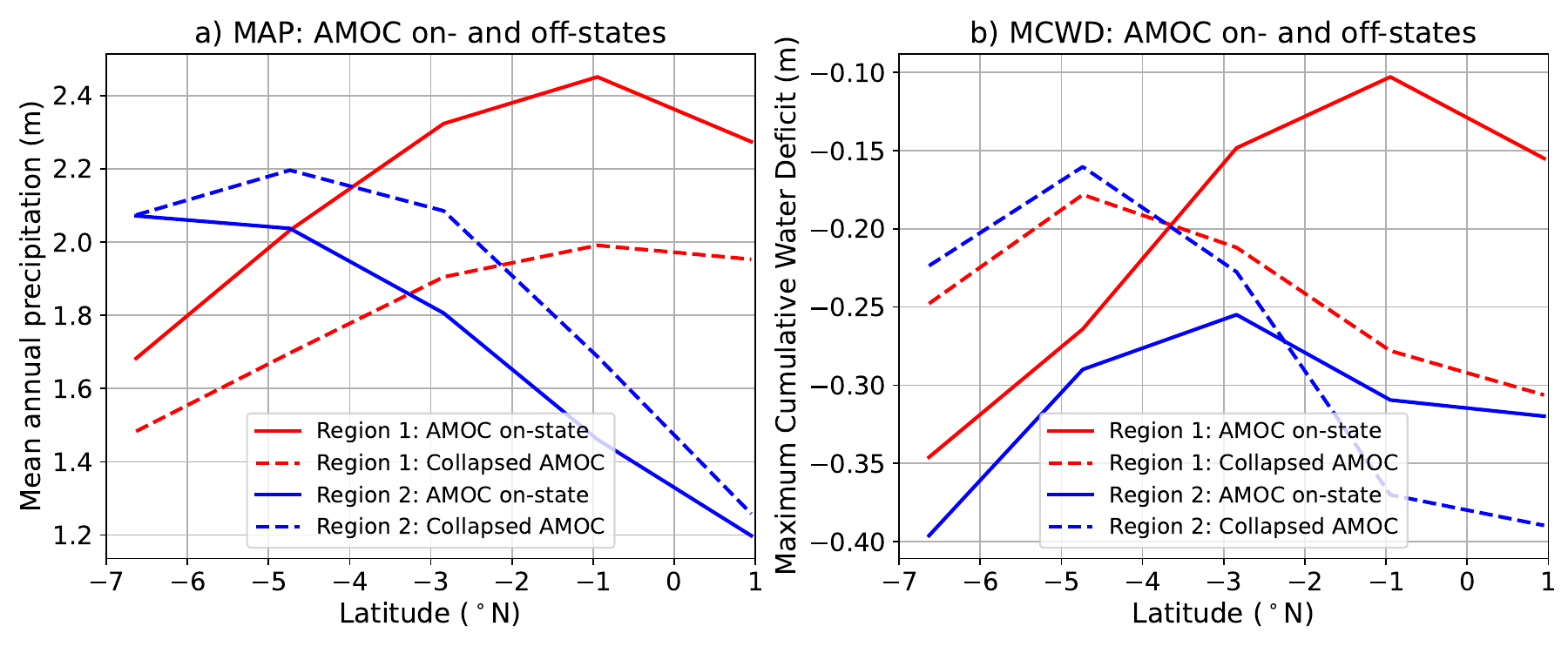}
	\caption{a) Mean Annual Precipitation (MAP) in the AMOC on-state, where $\Psi=\Psi_\mathrm{ON}$ (solid lines) and in the collapsed AMOC, where $\Psi=0$ (dashed lines) as a function of latitude in regions 1 (red lines) and 2 (blue lines).\\
		b) Maximum Cumulative Water Deficit in the AMOC on-state and in the collapsed AMOC state, for regions 1 and 2.}
	\label{fig:map_mcwd_states}
\end{figure}

\begin{figure}
	\centering
	\includegraphics[width=\textwidth]{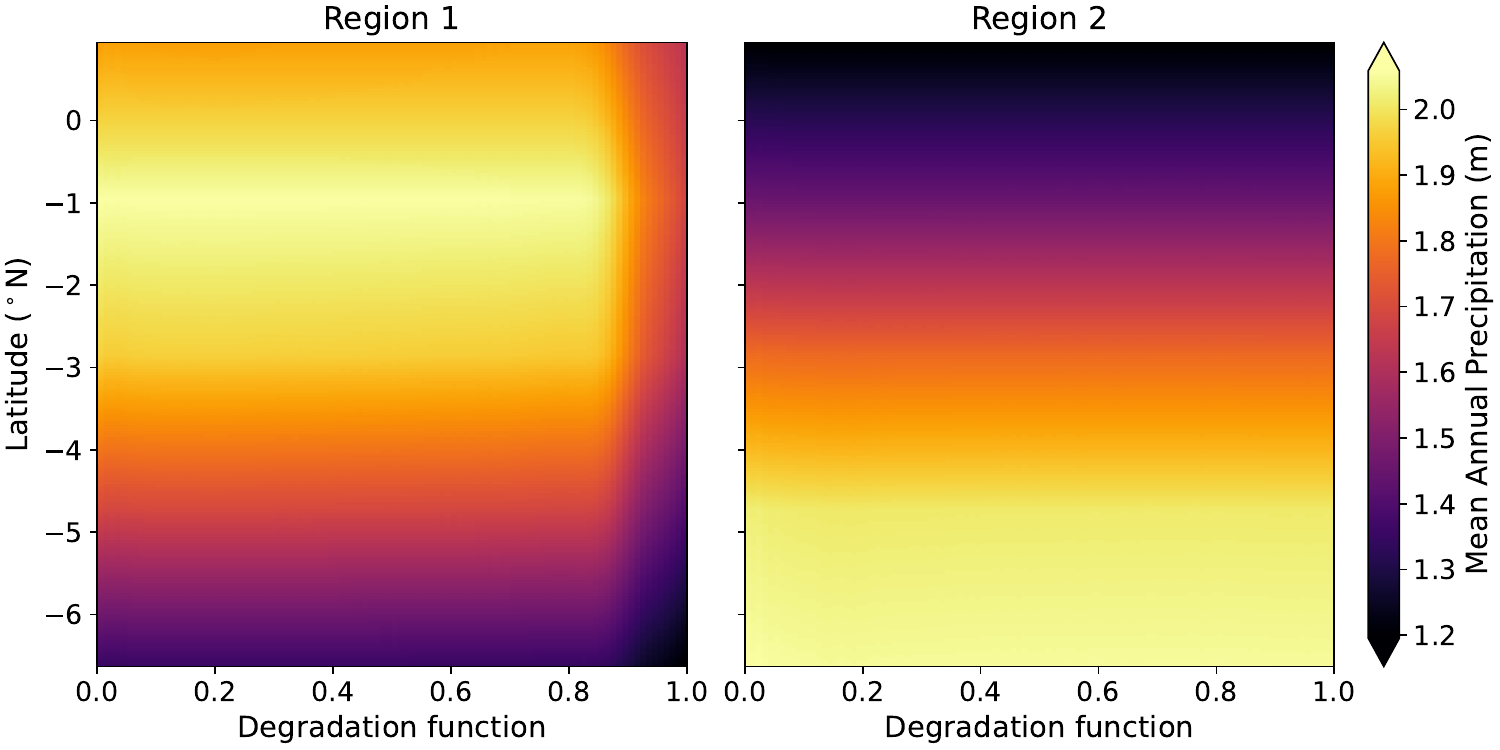}
	\caption{Mean Annual Precipitation (MAP) tracked across every level of the degradation function and rebuilt using the mean tracked AMOC strength $\Psi$ (see Fig.~\ref{fig:mean_amoc_strength}) for both regions shown in Fig.~\ref{fig:regions}.}
	\label{fig:rebuilt_map}
\end{figure}

\begin{figure}
	\centering
	\includegraphics[width=\textwidth]{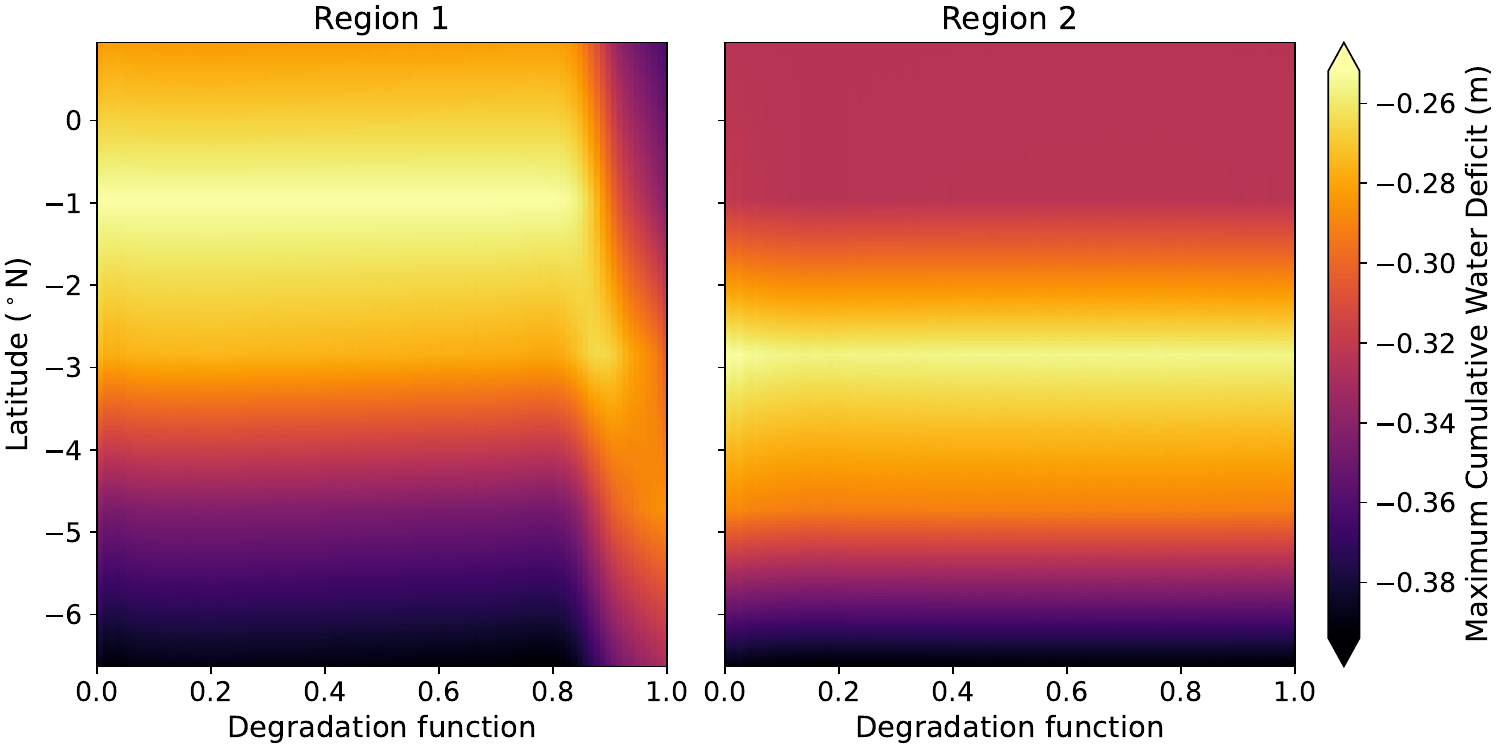}
	\caption{Maximum Cumulative Water Deficit (MCWD) tracked across every level of the degradation function and rebuilt using the mean tracked AMOC strength $\Psi$ (see Fig.~\ref{fig:mean_amoc_strength}) for both regions shown in Fig.~\ref{fig:regions}.}
	\label{fig:rebuilt_mcwd}
\end{figure}

\begin{figure}
	\centering
	\includegraphics[width=\textwidth]{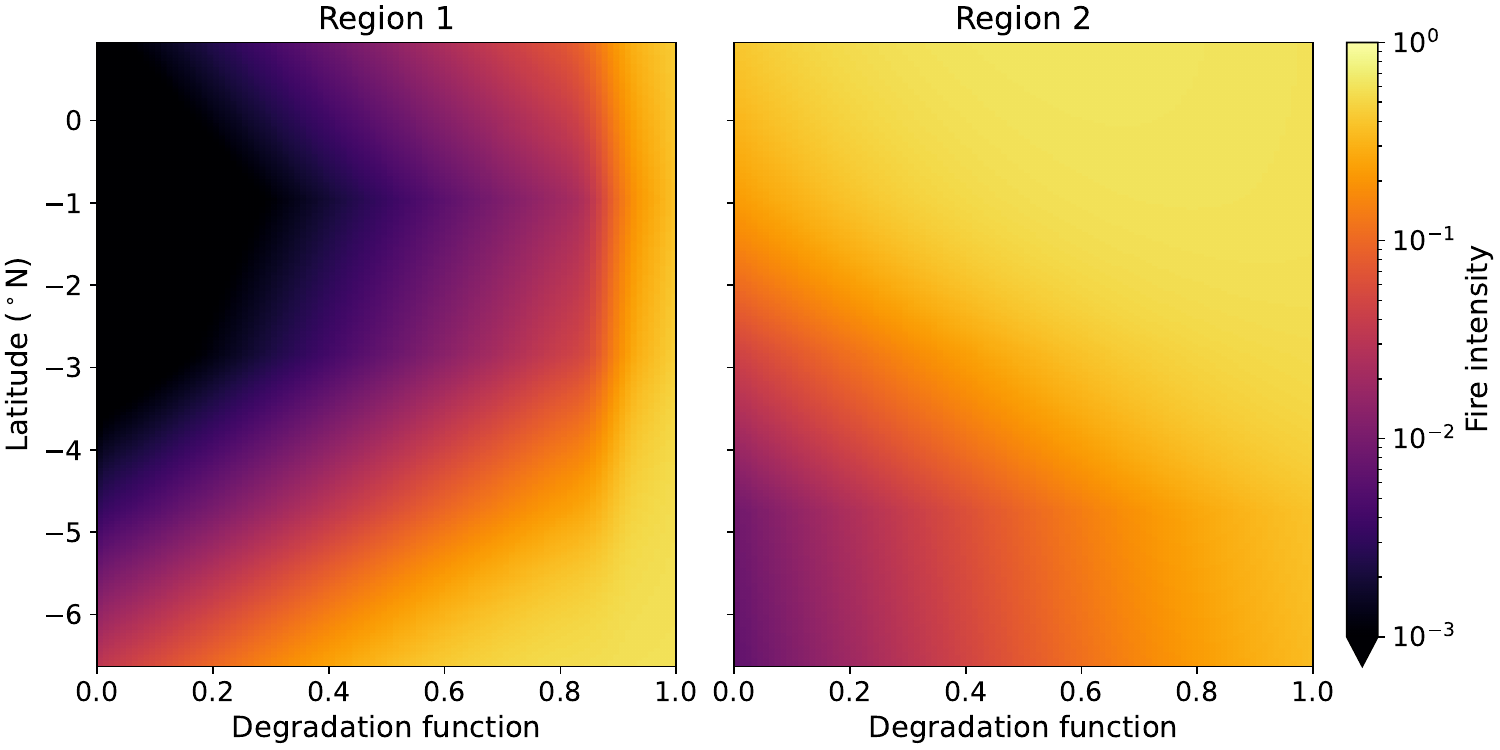}
	\caption{Mean fire intensity $f(T,\mathrm{MAP})$ (see Eq.~\ref{eq:fire_intensity}) tracked across every level of the degradation function for both regions shown in Fig.~\ref{fig:regions}. MAP was reconstructed from the average tracked AMOC strength $\Psi$ (see Fig.~\ref{fig:rebuilt_map}).}
	\label{fig:rebuilt_fire_intensity}
\end{figure}

\end{document}